\documentclass[aps,prd,amsfonts,nofootinbib,supercriptaddress]{revtex4}
\usepackage{amssymb,amsmath}
\usepackage{graphicx}
\usepackage{epsfig}
\usepackage{feynmp}

\newcommand{\beq}{\begin{equation}}
\newcommand{\eeq}{\end{equation}}
\newcommand{\bea}{\begin{eqnarray}}
\newcommand{\eea}{\end{eqnarray}}
\newcommand{\vc}[1]{{\textbf{#1}}}
\newcommand{\mc}[1]{\mathcal{#1}}

\begin{document}

\hfill HIP-2010-12/TH, ITP-UU-10-12, SPIN-10-10

\vspace{0.5cm}

\title{Path Integral for Inflationary Perturbations}

\author{Tomislav Prokopec$^{a}$  and Gerasimos Rigopoulos$^{b}$}
\affiliation{
$^a$Institute for Theoretical Physics and Spinoza Institute,
Utrecht University,
Leuvenlaan 4,
3584 CE Utrecht,
The Netherlands
\\$^b$Helsinki Institute of Physics, P.O.Box 64,
FIN-00014, University of Helsinki,
Finland
}

\begin{abstract} \noindent
The quantum theory of cosmological perturbations in single field inflation is formulated in terms of a path integral. Starting from a canonical formulation, we show how the free propagators can be obtained from the well known gauge-invariant quadratic action for scalar and tensor perturbations, and determine the interactions to arbitrary order. This approach does not require the explicit solution of the energy and momentum constraints, a novel feature which simplifies the determination of the interaction vertices. The constraints and the necessary imposition of gauge conditions is reflected in the appearance of various commuting and anti-commuting auxiliary fields in the action. These auxiliary fields are not propagating physical degrees of freedom but need to be included in internal lines and loops in a diagrammatic expansion. To illustrate the formalism we discuss the tree-level 3-point and 4-point functions of the inflaton perturbations, reproducing the results already obtained by the methods used in the current literature. Loop calculations are left for future work.
\end{abstract}

\maketitle

\section{Introduction}

Perhaps the most remarkable aspect of inflation~\cite{Guth:1980zm}
is its ability to imprint fluctuations on cosmic scales through a confluence
of quantum mechanics and general relativity. This connection, first realized more than 25 years ago \cite{Mukhanov:1981xt, Hawking:1982cz, Starobinsky:1982ee, Guth:1982ec, Bardeen:1983qw, Sasaki:1986hm}~\footnote{The first rigorous and quantitatively accurate treatments of inflationary perturbations were \cite{Mukhanov:1981xt} and \cite{Sasaki:1986hm}.}, has given inflationary theory the impetus which positioned it as the leading paradigm for approaching the physics of the early universe. Since then, the primordial fluctuations have been measured in the CMB~\cite{Bennett:1996ce} with ever increasing accuracy and resolution~\cite{Larson:2010gs, Komatsu:2010fb} and will be scrutinized even further in the near future~\cite{:2006uk}. It is not surprising then that over these past decades a lot of effort has been devoted to fleshing out the predictions inflation makes for these fluctuations in a variety of theoretical settings. Since these fluctuations are initially small, of order $10^{-5}$ at $z \simeq 1090$, the linearized theory of perturbations has been developed to a significant degree and has been used, rather successfully, to compare
theory to observation. Over the past few years, efforts have intensified to explore inflationary perturbations beyond linear order, mostly in the context of the related non-Gaussianity which has become a significant subfield of cosmological
research~\cite{Turner:2008zza}. The amount of work that has been done on
the subject is by now rather voluminous with many authors examining various aspects. A definitive calculation was performed by Maldacena~\cite{Maldacena:2002vr} showing that single field inflation leads to small primordial non-Gaussianities (see also \cite{Acquaviva:2002ud}) with more complicated
single- and multi-field models providing more possibilities for larger non-Gaussianity, see {\it e.g.} \cite{Bernardeau:2002jf, Zaldarriaga:2003my, Rigopoulos:2005ae, Rigopoulos:2005us, Vernizzi:2006ve, Alishahiha:2004eh, Enqvist:2004ey, Lyth:2005fi, Malik:2006pm, Sasaki:2006kq, Langlois:2008qf, Barnaby:2008fk, Chen:2008wn, Byrnes:2008wi, Sasaki:2008uc,Hotchkiss:2009pj, Chambers:2009ki}. On the observational side, a major effort is under way to develop observational measures of non-gaussianity \cite{Komatsu:2010fb, Creminelli:2005hu, Liguori:2010hx, Jeong:2009wi, Bucher:2009nm, Yadav:2007yy, Komatsu:2008hk, Smith:2009jr, Fergusson:2008ra, Seljak:2008xr, Slosar:2008hx}.

The foray into non-linear corrections and the associated non-Gaussianity
has been motivated by a number of reasons. If one neglects the running
of the spectral indices, at linear order inflationary
theories predict four numbers: the amplitude and spectral index of scalar
and tensor perturbations, and a variety of different models
can coincide on these predictions.
However, non-Gaussianity can be rather discriminatory for
different models due to its much richer, and more complicated, structure.
As an example, the detection of a significant three-point function would immediately rule out single-field inflation,
as well as some simple multi-field generalizations, may favor alternative models, or provide evidence for the processes involved in heating up the universe after Inflation. Entwined with considerations of testing inflationary theory against observations and non-Gaussianity are considerations of theoretical understanding: calculating and controlling higher order quantum loop corrections to inflationary predictions, backreaction issues~\cite{Vilenkin:1982wt}
and various divergences which appear at higher orders of perturbation theory -- see for example~\cite{Tsamis:1996qm, Weinberg:2005vy, Tsamis:1993ub, Weinberg:2006ac, Sloth:2006az, Sloth:2006nu, Seery:2007we, Seery:2007wf, Bilandzic:2007nb, Janssen:2008dw, Janssen:2009nz, Campo:2009fx, Urakawa:2009my, Senatore:2009cf}.

So far, all the work on inflationary non-linear corrections has been done using the operator language in the interaction picture. However, in many branches of modern physics it is the path integral formulation for quantum mechanical systems which has proved quite useful. For example, it has been of paramount importance in understanding gauge theories and their experimental consequences for particle physics, the theoretical description of condensed matter systems and is the preferred language in which modern theories of fundamental physics are formulated and quantized. In this spirit, and hoping to throw more light on the understanding of inflation, here we develop a path integral formulation for inflationary perturbations to arbitrary order in interaction terms. An early path-integral formulation of linear perturbation theory can be found in \cite{Anderegg:1994xq}. A novel feature of our approach is that there is no need to explicitly solve the energy and momentum constraints to a particular perturbative order as is usually done when working in a particular gauge. This allows us to obtain the interaction terms, and hence the vertices, to arbitrary order in a closed form.

We start from the more fundamental canonical path integral and show how to obtain the configuration phase space path integral, making in the process a connection with the well known gauge invariant linear perturbation theory. With a single inflaton there is only one scalar -- expressed in terms of the the Sasaki-Mukhanov variable -- and two tensor degrees of freedom which propagate, as expected,
while the vectors completely drop out at quadratic order. However, various (real and commuting) auxiliary fields which do not appear in
the in-state, the external lines in the ``in-in diagrammatic'' expansion, must be included in computations of N-point functions with $N\geq 4$ or calculations involving loops. These fields arise because the constraints are not explicitly solved. Furthermore, anti-commuting ghosts arise from the path integral measure which must also be taken into account at a certain loop order.
Here we focus on standard potential-energy-dominated single-field inflation, but the generalization to more fields and more complex theories is in principle straightforward as long as a canonical formulation is available.

The paper is organized as follows: In section \ref{Quantization of Inflationary perturbations} we derive the action for perturbations in a form that includes interactions to all orders in a closed expression, discuss the role of constraints and gauge conditions and formulate the transition amplitude between two quantum states separated by (background) time t in terms of a path integral. In the process the action is written in a form that makes contact with known results from gauge-invariant linear perturbation theory \cite{Mukhanov:1990me} and simplifies the computation of the propagators. In section \ref{The functional in-in formalism and expectation values} we discuss the ``in-in'' generating functional  which provides the appropriate diagrammatic rules for the computation of the  N-point functions relevant to cosmology. In section \ref{The 3-point and 4-point functions of Inflaton perturbations} we apply the above formalism to give expressions for the tree level 3-point and 4-point functions in terms of propagators, reproducing the results already obtained using an operator formalism and a different methodology. We close in section \ref{Discussion} with a discussion of our results.

\section{Quantization of Inflationary perturbations}
\label{Quantization of Inflationary perturbations}

We will consider a single scalar field $\chi$ minimally coupled to Einstein Gravity. The action  in the canonical form~\cite{Arnowitt:1962hi}
is ($\kappa^2 = 16\pi G_N = 1$)
\beq
S=\int d^4x\, \{ p^{ij}\partial_t g_{ij} +  p_\chi\partial_t \chi
           - N\mc{H} - N_i\mc{H}^i \}
\,,
\label{action1}
\eeq
with the hamiltonian $\mc{H}$ and momentum density $\mc{H}^i$
\beq
\mc{H}=\frac{1}{\sqrt{g}}\left(p^{ij}g_{ik}g_{jl}p^{kl}-\frac{1}{2}p^2\right)
   - \sqrt{g}R +\frac{1}{2}\frac{p_\chi^2}{\sqrt{g}}
   + \frac{1}{2}\sqrt{g}g^{ij}\nabla_i\chi\nabla_j\chi + \sqrt{g}V(\chi)
\,,
\label{ADM:hamiltonian}
\eeq
\beq
\mc{H}^i=p_\chi\partial^i\chi - 2\nabla_{\!\!j} \, p^{ij}\,,
\label{Momentum:ADM}
\eeq
where $p^{ij}$ and $p_\chi$ are the canonical momenta of the spatial metric $g_{ij}$
and scalar field $\chi$. They are tensor densities of weight one, i.e. $p^{ij}/\sqrt{g}$ and $p_{\chi}/\sqrt{g}$ are true tensors. In particular, the covariant derivative in~(\ref{Momentum:ADM})
is to be understood as,
$\nabla_{\!j} \, p^{ij} = \partial_j p^{ij} + \Gamma^i_{jl} p^{jl}$,
where $\Gamma^i_{jl}$ is the Levi-Civit\`a connection. Finally, $N$ and $N^i$ are the lapse and shift
functions, related to the temporal components of the metric tensor as
$g_{0i}=g_{ij}N^j$ and $g_{00} = -N^2 + N^ig_{ij}N^j$,
$R$ denotes the spatial Ricci scalar, $g={\rm det}[g_{ij}]$
and $p\equiv g_{ij}p^{ij}$. We will be interested in quantizing perturbations around a classical inflationary background, so let us set
\bea\label{perts1}
p^{ij} &=& \frac{\mathcal{P}(t)}{6a(t)}\left(\delta^{ij}+\pi^{ij}(t,\vc{x})\right) \,, \\ \label{perts2}
p_\chi &=& \mathcal{P}_\phi(1+\pi_\varphi(t,\vc{x}))\,,\\
g_{ij} &=& a(t)^2(\delta_{ij} + h_{ij}(t,\vc{x})) \,,\\ \label{perts3}
\chi &=&{\phi}(t)+\varphi(t,\vc{x})\,, \\
N &=& \bar{N}(t)+n(t,\vc{x})
\,.
\label{perts4}
\eea
Note that the background value of the shift $N_i$ is zero. The action $S^{(0)}$ for the background dynamics is obtained by setting
the perturbations to zero
\beq
S^{(0)} = \int d^3x dt \,\Big\{\mathcal{P}\frac{da}{dt}
        + \mathcal{P}_\phi\frac{d{\phi}}{dt}-\bar{N}\mathcal{H}_0(t)
\Big\}
\,,
\eeq
where
\beq
\mathcal{H}_0(t)
 = -\frac{\mathcal{P}^2}{24a}
            +\frac{\mathcal{P}_\phi^2}{2a^3}+a^3{V}
\,.
\eeq
The background equations of motion are obtained from $S^{(0)}$ by
varying with respect to $a$, $\phi$, $\mathcal{P}$ and $\mathcal{P}_\phi$, resulting in
\bea
\dot a&=&-\frac{\mathcal{P}}{12a}\,,
\label{dot a}
\\
\dot{\phi}&=&\frac{\mathcal{P}_\phi}{a^3}\,,\\
\dot{\mathcal{P}}&=&-\frac{\mathcal{P}^2}{24{a^2}}
                  +\frac{3}{2}\frac{\mathcal{P}_\phi^2}{a^4}-3a^2V\,,
\label{Pdot}\\
\dot{\mathcal{P}}_\phi  &=&-a^3V_{,\phi}\,,
\label{Pphidot}
\eea
where
$V = V(\phi(t))$, $V_{,\phi}=dV(\phi)/d\phi$
and we introduced the `dotted' derivative,
$ \dot a \equiv \bar N^{-1}da/dt$, {\it etc}.
Furthermore,
variation with respect to $\bar{N}$ gives the constraint $\mathcal{H}_0(t)=0$,
or equivalently,
\beq
\frac{\mathcal{P}^2}{24a}=\frac{\mathcal{P}_\phi^2}{2a^3}+a^3V
\label{Nbar:constraint}
\,.
\eeq
The well known equations of flat Friedmann-Lema\^itre-Robertson-Walker (FLRW)
cosmology are obtained with the identification
\beq
H \equiv \frac{\dot a}{a} = -\frac{\mathcal{P}}{12a^2}
\,.
\eeq
Indeed, inserting the solutions for the momenta $\mathcal{P}$ and
$\mathcal{P}_\phi$ into (\ref{Pdot}--\ref{Nbar:constraint}) one obtains,
\begin{eqnarray}
 && H^2 = \frac{\rho_\phi}{6}
         \equiv \frac{1}{6}\left(\frac{\dot \phi^2}{2} + V\right)
\label{background1}\\
&&  \dot H  = - \frac{\dot \phi^2}{4}
\label{background2}\\
&&  \ddot \phi + 3H\dot \phi + V_{,\phi} =  0
\label{background3}\\
&& \dot \rho_\phi + 3H (\rho_\phi + p_\phi)  = 0
\,,\qquad \Big(p_\phi = \frac{\dot \phi^2}{2}-V \Big)
\,.
\label{background4}
\end{eqnarray}
Recalling that a {\it dot} refers to $d/(\bar N dt)$,
we see that these equations
are time reparametrization invariant. For example, their form in
cosmological and conformal time is obtained simply by choosing the lapse
$\bar N=1$ and $\bar N = a$, respectively.
Of course, only two out of these four equations are independent. Indeed,
from the first equation and any one of the other three,
one can derive the remaining two equations.
To recover the dependence on the Newton constant $G_N$
one should reinsert $\kappa^2 = 16\pi G_N = 1$ into the {\it r.h.s.} of
the first two background equations~(\ref{background1}) and (\ref{background2}).

In such a model, inflation takes place when the field is rolling slowly on the slope of its potential. We therefore recall the `slow roll' parameters, which for the purpose of this
paper we define as,
\begin{equation}
  \epsilon \equiv -\frac{\dot{H}}{H^2}
              = \frac{3}{2}\frac{\dot\phi^2}{\rho_\phi}
\,,\qquad
\eta =  -\frac{\ddot\phi}{H\dot{\phi}}
\,.
\end{equation}
Notice that in the slow roll approximation when $\epsilon\ll 1$ and
$\eta\ll 1$, $\epsilon$ and $\eta$ reduce to the standard slow roll
definitions, $\epsilon\rightarrow (V_{,\phi}/V)^2$ and
$\eta\rightarrow (V_{,\phi\phi}/V)-\epsilon$.
For the rest of the paper we will be using $H$ instead of $\mc{P}$ and
$\mc{P}_\phi$ instead of $\dot{\phi}$.

\subsection{The action for perturbations}
\label{The action for perturbations}

We can now proceed to obtain the action for the perturbations by making the
replacements (\ref{perts1}--\ref{perts4}) and expanding~(\ref{action1}). We ignore terms linear in perturbations since they are multiplied by the background equations of motion (\ref{background1}--\ref{background4}), which we assume to hold. We can therefore write the complete action for the perturbations in canonical form as
\bea\label{action pert}
{ {\cal S}}_{\rm pert} &=&
 \int d^3xdt \;\Big( \mc{P}_\phi\pi_\varphi \partial_t\varphi
    -2a^3\!H\pi^{ij}\partial_t h_{ij}
  - \mc{H}_{\rm pert} + nC_0 + N_i C_i\Big)
    \equiv { {\cal S}}_{\rm F}+{ {\cal S}}_{\rm I}
\\
{ {\cal S}}_{\rm F} &=& \int d^3xdt\,\Big\{\mc{P}_\phi\pi_\varphi\partial_t\varphi
   - 2a^3H\pi^{ij}\partial_t h_{ij}
   -\mc{H}_{\rm F} + nC^{(1)}_0 + N_i C^{(1)}_i\Big\}
\label{S_F}
\\
{ {\cal S}}_{\rm I} &=& \int d^3xdt\,
          \Big\{-\mc{H}_{\rm I} + nC^{\geq 2}_0 + N_i C^{\geq 2}_i\Big\}
\,,
\label{S_I}
\eea
which completely specifies the dynamics of perturbations to all orders
in the interaction terms. For later convenience we have split the full action
${{\cal S}}_{\rm pert}$~(\ref{action pert})
into the free (quadratic) part ${{\cal S}}_{\rm F}$~(\ref{S_F})
and the interactions ${{\cal S}}_{\rm I}$~(\ref{S_I}), which include cubic,
quartic and other higher order terms in perturbations.
$C_0 = C_0^{(1)} + C_0^{\geq 2}
\equiv -\mc{H}^{(1)}/\bar N-\mc{H}_{\rm pert}/\bar N$
stands for the hamiltonian constraint, which is split into the linear part
(denoted by the superscript (1)) and quadratic and higher order parts
(denoted by the superscript $\geq 2$ and the subscript pert).
We have also split the momentum constraint as,
$C_i = C_i^{(1)} + C_i^{\geq 2}\equiv - \mc{H}^i$. Analogously,
$\mc{H}_{\rm pert}=\mc{H}_{\rm F}+\mc{H}_{\rm I}$ is the full
hamiltonian dictating the time evolution of the perturbations, split in Eqs.~(\ref{S_F}--\ref{S_I}) into a free
(quadratic) and an interacting part (qubic, quartic, etc). The Poisson bracket algebra of the system of perturbations can be calculated
to close
\bea
\{\mc{H}_{\rm pert}(\vc{x}),\,C_0(\vc{y})\}&=&-\left( C_i(\vc{x})\partial_i^y- C_i(\vc{y})\partial_i^x\right)\delta(\vc{x}-\vc{y})+ \frac{\partial_t C_0}{\bar{N}}\delta(\vc{x}-\vc{y})\,,\label{algebra1}\\
\{\mc{H}_{\rm pert}(\vc{x}),\,C_i(\vc{y})\}&=&C_0(\vc{y})\partial_i^x \delta(\vc{x}-\vc{y}) +\frac{\partial_t C_i}{\bar{N}}\delta(\vc{x}-\vc{y})\,,\label{algebra2}\\
\{C_{0}(\vc{x}),\,C_0(\vc{y})\}&=&\left( C_i(\vc{x})\partial_i^y- C_i(\vc{y})\partial_i^x\right)\delta(\vc{x}-\vc{y})\,,\label{algebra3}\\
\{C_{i}(\vc{x}),\,C_0(\vc{y})\}&=&C_0(\vc{x})\partial_i^y\delta(\vc{x}-\vc{y})\,,\label{algebra4}\\
\{C_{i}(\vc{x}),\,C_j(\vc{y})\}&=&\left( C_j(\vc{x})\partial_i^y- C_i(\vc{x})\partial_j^y\right)\delta(\vc{x}-\vc{y})\label{algebra5}\,,
\eea
where the fundamental Poisson brackets are seen from (\ref{action pert}) to be
\bea
\{\varphi(\vc{y},t),\,\pi_\varphi(\vc{x},t)\}
 &=&\frac{1}{\mc{P}_\phi}\,\delta(\vc{x}-\vc{y})\,,
\\
\{h_{kl}(\vc{y},t),\,\pi^{ij}(\vc{x},t)\}&=&
-\frac{1}{4a^3H}\left(\delta_{ik}\delta_{jl}
 +\delta_{il}\delta_{jk}\right)\delta(\vc{x}-\vc{y})\,.
\eea
We see that from a canonical point of view, inflationary perturbations form a \emph{constrained Hamiltonian system} \cite{Dirac,Faddeev:1969su}\footnote{For an early treatment of inflationary perturbations as a constrained Hamiltonian system see \cite{Makino:1991sg}.}.

What is the number of dynamical degrees of freedom in such a system? Following the point of view in \cite{Faddeev:1969su}, we see that the seven fields $(\varphi,h_{ij})$ appearing in~(\ref{action pert})
are not all independent degrees of freedom: $n$ and $N_i$ are lagrange multipliers without dynamics of their own, imposing under variation the conditions $C_0=0$ and $C_i=0$ which constrain the evolution of the system to take place in a lower dimensional hypersurface of the full 14-dimensional
phase space. The functions $n$ and $N_i$ are not determined by the dynamics and are arbitrary; in this particular case they parameterize the freedom in choosing how to break
spacetime into spatial hypersurfaces and time. Thus, the dynamical equations contain four completely free functions
which need to be fixed by the imposition of four further \emph{gauge conditions}: $\mc{Q}_\alpha(h_{ij},\varphi,\pi^{ij},\pi_\varphi)=0$. These gauge conditions can be chosen at will, up to the requirement
\beq\label{admissible1}
\{\mc{Q}_\alpha,\,\mc{Q}_\beta\}=0 \,,\quad
{\rm Det}\left\{\mc{Q}_\alpha,C_\beta\right\}\neq0\,.
\eeq
We see that the physical phase space has dimension $14-4-4=6$.
Indeed, as is well known and will be re-derived below,
there is only one dynamical scalar and a transverse traceless tensor
propagating in single field inflation, corresponding to 3 degrees of freedom
and a 6 dimensional dynamical phase space. Note that once a choice of $\mc{Q}_\alpha$ has been made, the lagrange multipliers $N_\alpha\equiv\{n,N_i\}$ can be determined by the imposition of the consistency relation
\beq\label{consistency}
\dot{\mc{Q}}_\alpha=\{\mc{H}_{\rm pert},\mc{Q}_\alpha\}-N_\beta\{C_\beta,\mc{Q}_\alpha\}\thickapprox 0\,,
\eeq
where the symbol $\thickapprox$ implies that the constraints $C_\beta=0$ are imposed \emph{after} the poisson brackets are evaluated. Condition (\ref{admissible1}) then insures that equation (\ref{consistency}) can be solved for $N_\alpha$.

Since the number of dynamical fields in the system is much less than the apparent dimension of its phase space, it proves convenient to explicitly separate out the 3 propagating degrees of freedom from the rest. We do this in the next paragraph, leading to equation (\ref{2nd-w-h}). Let us note that the physical degrees of freedom could be isolated by actually solving the constraints, thus expressing some of the canonical variables in terms of the others, and plugging the solution back in the action. This is feasible for the free theory, where the
interactions ${ {\cal S}}_{\rm I}$ in~(\ref{S_I}) are
ignored~\cite{Anderegg:1994xq}.
If the interactions are included however, the solution of the constraints can only be found by an iteration procedure and leads to rather cumbersome expressions. The approach we follow in this paper dispenses with the need to explicitly solve the constraints, although by direct comparison it is equivalent to such a solution at the linear level.

Recall now Eqs.~(\ref{action pert}--\ref{S_I}), where
the action for perturbations was split into the free and interacting
parts. Explicating the free action ${ {\cal S}}_{\rm F}$~(\ref{S_F}),
the free hamiltonian reads,
\bea
\mc{H}_{\rm F}
 &=&
4\bar{N}a^3H^2\left[\frac{1}{2}\pi^{ij}A_{ijkl}\pi^{kl}
          + \pi^{ij}\left(h_{ij} -\frac{1}{2}\delta_{ij}h\right)\right]
 + \bar{N}\frac{\mc{P}_\phi^2}{2a^3}\left(\pi_\varphi^2-h\pi_\varphi\right)
 + \bar{N}\frac{a}{2}(\partial_i\varphi)^2
 + \bar{N}\frac{a^3}{2}\Big[V_{,\phi\phi}\varphi^2 + V_{,\phi}\varphi h\Big]
\nonumber\\
&&
 + \,\,\bar{N}\left(4a^3H^2
                  + \frac{\mc{P}_\phi^2}{a^3}
              \right)
     \frac{h_{ij}h_{ij}}{4}
+ \,\,\bar{N}a\left(\frac{1}{4} h \nabla^2 h
                      - \frac{1}{2}h\partial_i\partial_j h_{ij}
                      + \frac{1}{2}h_{ij} \partial_i\partial_l h_{jl}
                      - \frac{1}{4} h_{jl}\nabla^2 h_{jl} \right)
\,,
\label{perturbation hamiltonian}
\eea
where we have used the notation
$A_{ijkl}\equiv \delta_{ik}\delta_{jl}+\delta_{il}\delta_{jk}
 -\delta_{ij}\delta_{kl}$, and
\bea
C^{(1)}_0 &=& a^3\left(\frac{\mc{P}_\phi^2}{2a^6}-2H^2\right)h
 + a\left(\partial_i \partial_jh_{ij} -\nabla^2 h\right)
 - a^3V_{,\phi}\varphi - \frac{\mc{P}_\phi^2}{a^3}\pi_\varphi
  + 4a^3H^2\pi \,,
\label{C(1)_0}
\\
C^{(1)}_i &=& -\frac{\mc{P}_\phi}{a^2}\partial_i\varphi
    -4aH \Big(\partial_l\pi^{li}
               + \partial_lh_{li} + \frac12\partial_i h
            \Big)\,,
\label{C(1)_i}
\eea
and where $\nabla^2 = \delta_{ij}\partial_i\partial_j$.
From now on a superscript/subscript $\geq n$ will indicate that
only terms of order $\geq n$ in the perturbations are included, such
that $C_\mu^{\geq 2}$ denotes contributions that are quadratic or higher
order in the perturbations.

Next, it will be convenient to use the standard scalar-vector-tensor
decomposition of the spatial metric perturbation,
\begin{equation}
 h_{ij} = \frac{\delta_{ij}}{3}h
     + \Big(\partial_i\partial_j-\frac{\delta_{ij}}{3}\nabla^2\Big)\tilde h
       + \partial_{(i}h^T_{j)}  + h^{TT}_{ij}
\label{hij:SVT}
\end{equation}
with
\begin{equation}
\partial_{(i}h^T_{j)}=\frac{1}{2}(\partial_ih^T_j+\partial_jh^T_i)
\,, \qquad
\partial_i h^T_i = 0
\,,\qquad
 \partial_i h^{TT}_{ij} = 0 = \partial_j h^{TT}_{ij}
\,.
\label{hij:SVT2}
\end{equation}
Furthermore, we will decompose the shift into the longitudinal and transverse
components as,
\beq
N_i=\partial_iS + N_i^T
\,,\qquad
{\rm with}\;\;\;\;
\partial_iN_i^T=0
\,.
\eeq
After a series of manipulations, which we present in appendix B,
the free action ${\cal S}_{\rm F}$ can be diagonalized.
In summary, 
if we define
\begin{equation}
   w = (h-\nabla^2 \tilde h) - \frac{3}{\sqrt{\epsilon}}\varphi\,,
\label{scalar variable:w}
\end{equation}
and perform the following linear shifts in all the perturbation fields:
\bea
n&=&\tilde{n}+L_0
\label{shift-n}\\
N_i&=&\tilde{N}_i + L_i
\label{shift-Ni}\\
\pi^{ij}&=&\rho^{ij}+L_{ij}
\label{shift-rho}\\
\pi_\varphi&=&\rho_\varphi+L_\varphi
\label{shift-rhophi}
\,,
\eea
where $L_0$, $L_i$, $L_{ij}$ and $L_\varphi$ are given
in Eqs.~(\ref{L0}--\ref{appendix:J}) of appendix B,
we find that, up to boundary terms,  the free action 
takes the form
\begin{eqnarray}
 {\cal S}_{\rm F}  &=& \int d^3x\bar{N}dt a^3
       \bigg\{\frac{\epsilon}{18}
                          \Big[{\dot w}^2 -
\Big(\frac{\partial_iw}{a}\Big)^2\,
                          \Big]\,+\,\frac{1}{4}\bigg[(\dot
h^{TT}_{ij})^2
               -  \Big(\frac{\partial_l h^{TT}_{ij}}{a}\Big)^2\bigg]
\nonumber\\
&&\hskip 2cm -2H^2\epsilon\,{\rho}_\varphi^2
             - 2H^2{\rho}^{ij}A_{ijkl}{\rho}^{kl}
             -\frac{2(3-\epsilon)H^2}{\bar N^2}\tilde n^2
             - \frac{1}{2a^4\bar N^2}
                 \Big({\tilde N}^T_j\nabla^2{\tilde N}^T_j
                  - \frac{4}{3-\epsilon}(\nabla^2 {\tilde S})^2 \Big)
      \bigg\}
\label{2nd-w-h}
\,.
\end{eqnarray}

In this form, the action clearly shows that the perturbations which propagate are
one scalar degree of freedom $w$ and the 2 degrees of freedom of the transverse traceless tensor $h_{ij}^{TT}$. All the other fields are \emph{auxiliary} fields without dynamics of their own. The first line of (\ref{2nd-w-h}) is well known from the gauge invariant theory
of cosmological perturbations~\cite{Mukhanov:1990me}~\footnote{The scalar
perturbation is usually expressed in terms of the Sasaki-Mukhanov variable
$v = a(\varphi - z (h-\nabla^2 \tilde h))$ with $z={\cal P}_\phi/[6a^3H]$,
and the corresponding free action, when written in conformal time, is shown in
Eq.~(\ref{2nd-order:scalar:v}) of appendix B.}.
The fields $w$ and $h^{\rm TT}_{ij}$ in Eq.~(\ref{2nd-w-h}) are gauge invariant under
\emph{linear} gauge transformations and time independent on
long wavelengths~\footnote{If the gauge $\varphi=0$ is chosen,
$w$ represents the curvature perturbation and is also constant during inflation
on long wavelengths to all orders in perturbation
theory~\cite{Maldacena:2002vr}.}. It is interesting to note that all the auxiliary fields, both $\tilde{n}$,
$\tilde{N}^T_i$ and $\tilde{S}$ as well as the shifted momenta $\rho^{ij}$ and $\rho_\varphi$ are also invariant under such linearized transformations -- see Eq.~(\ref{transformation:pi_varphi+piij}) in appendix B and the discussion that precedes it.

The variation of the action~(\ref{2nd-w-h})
{\it w.r.t.} $\pi_\varphi$, $\pi^{ij}$, $\tilde n$, $\tilde N^T_j$
and $\tilde S$ yields
the on-shell relations, $\rho_\varphi=0$, $\rho^{ij}=0$, $\tilde n=0$,
$\tilde N^T_j=0$ and $\tilde S=0$.
The relations $\rho_\varphi=0$ and $\rho^{ij}=0$ translate into
the relation between the momenta $\pi^{ij}$ and $\pi_\varphi$
and the time derivatives of $h_{ij}$ and $\varphi$.
On the other hand, $\tilde{n}=0$, $\nabla^2\tilde{N}^T_j=0$
and $\nabla^2\tilde{S}=0$ translate into the solutions of
the linear constraints. From Eqs.~(\ref{L0}--\ref{Li}) and~(\ref{appendix:J})
we see that the lapse perturbation $n$ and the shift $N_i$
are then given in terms of the fields $(\varphi,h_{ij})$
and their one time derivative or one or two spatial derivatives.
Thus, the free action~(\ref{2nd-w-h}) contains all the elements of the
well known gauge invariant treatment of linear inflationary perturbations
and the transition between the hamiltonian and lagrangian formulation
of the problem.

Interactions are encoded in ${\cal S}_{\rm I}$ given in
Eq.~(\ref{S_I}), with
\beq
C_0^{\geq 2} = - \frac{1}{\bar{N}}\left(\mc{H}_{\rm F}+\mc{H}_{\rm I}\right)
\label{C0gt2}
\eeq
and
\beq
C_i^{\geq 2} = -\, \frac{\mc{P}_\phi}{a^2}(\tilde{g}^{il})_{\geq 1}\partial_l\varphi
 - \frac{\mc{P}_\phi}{a^2}\pi_\varphi\tilde{g}^{il}\partial_l\varphi
 -4aH\left\{
       (\tilde{g}^{il})_{\geq 1}
                   \Big(\partial_jh_{jl}-\frac12\partial_lh\Big)
         + \tilde{g}^{il} \Big(\partial_jh_{lr}-\frac12\partial_lh_{jr}\Big)
                  \pi^{jr}
     \right\}
\,,
\label{Cigt2}
\eeq
where $\mc{H}_{\rm F}$ and $\mc{H}_{\rm I}$ are given in
Eq.~(\ref{perturbation hamiltonian}) and Eq.~(\ref{interaction hamiltonian})
of appendix D, respectively. After all the variables
are expressed in terms of the redefined fields,
$\tilde{n}$, $\tilde{N}_i$, $\rho^{ij}$ and $\rho_\varphi$
it becomes clear that ${\cal S}_I$ cannot be written solely in terms of $w$, $h_{ij}^{\rm TT}$ and the auxiliary fields which we will collectively call $\sigma$. Indeed, if $\varphi$ is exchanged for $w$ in ${\cal S}_I$, $ \varphi \rightarrow \frac{\sqrt{\epsilon}}{3}(h-\nabla^2\tilde{h}-w)$, one obtains
an interaction action~(\ref{S_I}) which is a functional not only of
$w,\,h^{TT}_{ij}$ and $\sigma$, but also of $h,\,\tilde{h}$ and $h^{\rm T}_i$
(for a more explicit form of ${\cal S}_{\rm I}$ see
Eqs.~(\ref{interaction hamiltonian}) and~(\ref{C0gt2}--\ref{Cigt2})).
Thus, the interaction terms seem to contain four more fields than those appearing in ${\cal S}_F$~(\ref{2nd-w-h}) which determines the free dynamics. This is of course related to the aforementioned arbitrariness in the time evolution of the system. The imposition of the four gauge conditions $\mc{Q}_\alpha=0$ cures this arbitrariness by expressing $h$, $\tilde{h}$ and $h^{\rm T}_i$ in terms of the fields appearing in ${\cal S}_{\rm F}$. This observation also demonstrates the necessity to consider \emph{non-linearized} gauge transformations when interactions are included. Notice that ${\cal S}_F+{\cal S}_{\rm I}$ is explicitly invariant under linearized transformations, and hence it would seem that the action as a whole is not since ${\cal S}_I$ explicitly depends on the gauge chosen, even at the linear level. This contradicts the well known exact gauge invariance of the action\footnote{Up to boundary terms which we ignore in this paper.}. The resolution of this seeming paradox lies in the observation that the exact gauge invariance of the action implies that for interaction terms of order up to $n$, one \emph{must be} able to absorb any such apparently gauge dependent terms by non-linear \emph{field redefinitions}: gauge transformations of order $n-1$. An explicit example for cubic interactions has been worked out in \cite{Maldacena:2002vr}.
With this observation, quantities that have been calculated in different gauges can be compared using the appropriate field redefinitions.

\subsection{The path integral}
\label{The path integral}

The results of the previous section can be used to obtain a convenient
quantization scheme for the perturbations, including the interactions.
According to the above discussion, inflationary perturbations are described
by a constrained hamiltonian system with action (\ref{action pert}).
In general, the quantization of such systems is easily formulated in terms
of a path integral~\cite{Faddeev:1969su}. Indeed, the transition amplitude
from state $|\mc{A};t_\mc{A}\rangle$ at $t=t_\mc{A}$ to $|\mc{B}; t_\mc{B}\rangle$ at time $t_\mc{B}$
can be written as
\beq
\langle\mc{B}; t_{\mc{B}}| \mc{A};t_\mc{A}\rangle \,
= \int \limits_{(h_{ij},\varphi)(t_\mc{A})=\mc{A}}^{(h_{ij},\varphi)(t_\mc{B})=\mc{B}}
\!\!\!\![{\cal D}h_{ij}{\cal D}\varphi {\cal D}\pi^{ij}
 {\cal D}\pi_\varphi {\cal D}n {\cal D}N_i]\,
 \prod\limits_\alpha\delta\left[\mc{Q}_\alpha\right]\,
\left|{\rm Det}\left\{\mc{Q}_\alpha,C_\beta\right\}\right|\,
{\rm e}^{\,i{\cal S}_{\rm pert}}
\,,
\label{transition Amplitude 1}
\eeq
where ${\cal S}_{\rm pert}$ is given in (\ref{action pert}--\ref{S_I}), integrated over time
$t^\prime\in[t_\mc{A},t_\mc{B}]$,
and we have schematically denoted the initial ($\mc{A}$) and final ($\mc{B}$)
field configurations by the limits of the path integral.
The transition amplitude~(\ref{transition Amplitude 1}) is
the fundamental quantity for calculating quantum correlators of
any kind and the starting point for calculating the correct lagrangian
path integral. According to \cite{Faddeev:1969su}, it is \emph{independent} of the choice of gauge conditions $\mc{Q}_\alpha$, as long as the boundary states are appropriately defined in an invariant manner. We shall address this issue
in a forthcoming publication.
Note the appearance of a non-trivial measure containing
$\left|{\rm Det}\left\{\mc{Q}_\alpha,C_\beta\right\}\right|$,
the determinant of the Poisson brackets between the constraints
$C_\alpha$ and the gauge conditions $\mc{Q}_\alpha$.
Equation~(\ref{transition Amplitude 1}) is a self-contained starting point
for doing perturbative calculations.
By splitting the action as above in~(\ref{action pert}--\ref{S_I})
\beq
{\cal S}_{\rm pert}={\cal S}_{\rm F} + {\cal S}_{\rm I}
\,,
\label{actin pert: split}
\eeq
it is possible to evaluate the path integral~(\ref{transition Amplitude 1})
perturbatively, expanding in powers of the interaction action,
after including an appropriate representation of the determinant.
The free action ${\cal S}_{\rm F}$ then determines the propagators while
${\cal S}_{\rm I}$ gives the various interaction vertices, completely
defining a diagrammatic expansion.

The focal point of the previous section has been the free
action~(\ref{S_F}) and~(\ref{perturbation hamiltonian}--\ref{C(1)_i})
that can be brought to the diagonal form (\ref{2nd-w-h})
from which propagators can be easily obtained.
The field redefinitions (\ref{shift-n}--\ref{shift-rhophi}) are
linear shifts in all the fields of the action which do not affect the form
of the functional measure in the path integral. More precisely, they contribute time dependent -- but field independent -- factors which cancel in our later computations.
The transition amplitude can therefore also be written in terms
of the redefined fields as
\beq\label{transition Amplitude2}
\langle\mc{B}; t_\mc{B}| \mc{A};t_\mc{A}\rangle \,
= \int \limits_{(h^{TT}_{ij}\!,\,w)(t_\mc{A})=\mc{A}}^{(h^{TT}_{ij}\!,\,w)(t_\mc{B})=\mc{B}}
\!\!\!\! [{\cal D}w{\cal D}h^{TT}_{ij}{\cal D}\sigma{\cal D}h{\cal D}\tilde{h}{\cal D}h^T_j ]\,
\prod\limits_\alpha\delta\left[\mc{Q}_\alpha\right]\,
\left|{\rm Det}\left\{\mc{Q}_\alpha,C_\beta\right\}\right|\,
{\rm e}^{\,i{\cal S}_{\rm F}[w,\,h^{TT}_{ij},\,\sigma]\,+\,i{\cal S}_{\rm I}},
\eeq
making it evident that the physical propagating degrees of freedom of the system are the quanta of $w$ and $h^{TT}_{ij}$. ${\cal D}\sigma \equiv
{\cal D}{\rho}^{ij}
{\cal D}{\rho}_\varphi{\cal D}{\tilde n}{\cal D}{\tilde N}^T_i
{\cal D}{\tilde S}$ collectively denotes integration over the 11 auxiliary fields. One can note that the full action is quadratic in $\rho_{ij}$ and $\rho_\varphi$ which could in principle be integrated out. However, it proves convenient to keep them
along with $\tilde{n}$, $\tilde{N}^T_i$ and $\tilde{S}$.
The ``propagation'' of all these auxiliary fields is determined by
the second line of (\ref{2nd-w-h}) and is trivial.
We here use the term propagation for the auxiliary
fields rather loosely since they are not attached to wave operators. Since $w$ and $h^{TT}_{ij}$ are the dynamical degrees of freedom, it is clear that the initial and final states should be determined in terms of their free quanta.

To complete the derivation of the more workable expression
for the path integral,
we are left with the task of evaluating the determinant
$\left|{\rm Det}\left\{\mc{Q}_\alpha,C_\beta\right\}\right|$
in~(\ref{transition Amplitude2}). Denoting
\beq
 \Omega_{\alpha\beta}
     \equiv \left\{\mc{Q}_\alpha,C_\beta\right\}\!\big|_{\mc{Q}_\gamma=0}
\,,
\label{ghost-operator}
\eeq
where the gauge condition ${\mc{Q}_\gamma=0}$
has been imposed after evaluating
the Poisson brackets, we can represent the determinant as a path integral
over anticommuting scalar fields -- Faddeev-Popov ghosts --
$\bar{\eta}_\alpha$ and $\eta_\alpha$~\footnote{See for example
Eqs.~(5.1.31-32) in Ref.~\cite{Ramond:1981pw}.}
\beq
\left|{\rm Det} \Omega_{\alpha\beta}\right|
 = \int [{\cal D}\bar{\eta}_\gamma{\cal D}\eta_\gamma] \,\,
{\rm e}^{\rm{i} \!\int{\rm d}^4\!x \,
   \bar{\eta}_\alpha \Omega_{\alpha\beta}\eta_\beta}
\,.
\eeq
The transition amplitude is then written as
\beq\label{transition Amplitude3}
\langle\mc{B}; t_{\mc{B}}| \mc{A};t_{\mc{A}}\rangle \,
= \int \limits_{(h^{TT}_{ij}\!,\,w)(t_\mc{B})=\mc{A}}^{(h^{TT}_{ij}\!,\,w)(t_\mc{B})=\mc{B}}
\!\!\!\! [{\cal D}w{\cal D}h^{TT}_{ij}{\cal D}\sigma {\cal D}h^T_j{\cal D}h{\cal D}\tilde{h}{\cal D}
\bar{\eta}{\cal D}\eta]\, \prod\limits_\alpha\delta\left[\mc{Q}_\alpha\right]\,
\, {\rm e}^{\,{\rm i}{\cal S}_{\rm F}\,+\,{\rm i}\!
 \int{\rm d}^4\!x \,\bar{\eta}_\alpha\Omega^{\rm F}_{\alpha\beta}\eta_\beta\,
+\,{\rm i}{\cal S}_{\rm I}\,
+\,{\rm i}\!\int{\rm d}^4\!x \,
       \bar{\eta}_\alpha\Omega^{\rm I}_{\alpha\beta}\eta_\beta}
\,,
\eeq
where we have split
$\Omega_{\alpha\beta}=\Omega^{\rm F}_{\alpha\beta}
                      + \Omega^{\rm I}_{\alpha\beta}$
into a free part,
determining the "propagation" of $\bar{\eta}_\alpha$ and $\eta_\alpha$,
and an interaction part which contains the other fields and couples them
to $\bar{\eta}_\alpha$ and $\eta_\alpha$. Two common gauge choices
$\mc{Q}_\alpha$ and the resulting matrices $\Omega_{\alpha\beta}$
are discussed in the appendix. We now have a complete description of the transition amplitude amenable to perturbative calculations.
Equation~(\ref{transition Amplitude3}) is the central result of this paper.

\section{The functional in-in formalism and expectation values}
\label{The functional in-in formalism and expectation values}

The previous section described the derivation of a path integral representation for the transition amplitude between states of inflationary perturbations in the past and the future, separated by time $t$. The perturbative evaluation of such amplitudes between the far past and the far future leads to the usual Feynman diagrams for the calculation of S-matrix elements, {\it i.e.} scattering amplitudes between in-states and out-states, useful in deriving predictions for the scattering experiments of particle physics. However, in cosmological applications the quantities of interest are not scattering amplitudes between predefined states (a boundary value problem), but expectation values of operators at a time t, given an initial state $ |\rm{in},t_\mc{A}\rangle$ at time $t_\mc{A}$ (an initial value problem). For the purpose of this work, this state is defined in terms of the freely propagating modes of $w$ and $h_{ij}^{\rm TT}$ at very early times: $t_\mc{A}\rightarrow t_{-\infty}$, and is taken to be an appropriately defined vacuum state $|\Omega,t_{-\infty}\rangle$. More generally, one has to begin the
evolution from a finite time and with an initial state chosen such that it
includes corrections due to the interactions, such that all field correlators
are finite at the initial hypersurface.
In this setting, the calculation of expectation values leads to a modification of the standard diagrammatic rules of QFT. Here, we sketch the functional in - in formalism for obtaining expectation values and arrive at the appropriate diagrammatic rules. For more details see \cite{Jordan:1986ug, Calzetta:1986ey} as well as the appendix of \cite{Weinberg:2005vy}.

\subsection{The in-in generating functional}

In analogy with standard QFT, one defines the in-in generating functional $Z[J_-,J_+]$ which contains \emph{two} arbitrary and independent sources $J_+$ and $J_-$
\beq\label{in-in}
Z[J_-,J_+]=\sum\limits_\alpha \langle \Omega,t_{-\infty}|\alpha,t_{out}\rangle_{J_-} \langle \alpha,t_{out}|\Omega,t_{-\infty}\rangle_{J_+}\,,
\eeq
where $\langle \alpha,t_{out}|\Omega ,t_{-\infty}\rangle_{J}$ denotes the transition amplitude from the aforementioned vacuum state at time $t_{-\infty}$ to a state $|\alpha,t_{out}\rangle$ at time $t_{out}$ in the presence of a source $J$. Expression (\ref{in-in}) can be interpreted as going forward in time in the presence of a source $J_+$ and then returning back in time in the presence of a source $J_-$. This explain the term ``closed-time-path formulation'' that is sometimes also used. However, in what follows such an interpretation is not needed. The final time $t_{out}$ can be taken to be any time later than all the times of interest in the calculation; here we take $t_{out}$ to be in the far future, $t_{out}\rightarrow +\infty$. Obviously, $Z[J,J]=1$.

The transition amplitudes appearing in (\ref{in-in}) can be expressed in terms of path integrals. Collectively denoting the relevant fields by $\xi$ we have
that
\beq
Z[J_-,J_+]=\int[{\cal D}\xi_+{\cal D}\xi_-] \,
{\rm e}^{iS[\xi_+]\,-\,iS[\xi_-]\,+\,iJ_+\xi_+\,-\,iJ_-\xi_-}
\,,
\eeq
with boundary conditions appropriate for the state $|\Omega,t_{-\infty}\rangle$ at $t_{-\infty}$ and the constraint that $\xi_+(t_{out})=\xi_-(t_{out})$.  Once $Z[J_-,J_+]$ has been calculated, expectation values can be obtained by taking variational derivatives:
\bea
&&\frac{\delta}{-i\delta J_-(y_1)}
\cdots \frac{\delta}{-i\delta J_-(y_m)}\frac{\delta}{i\delta J_+(x_1)} \cdots \frac{\delta}{i\delta J_+(x_n)}
Z[J_-,J_+]\Big|_{J_+=J_-=0}
\label{variational-derivs}
\\
&&\hskip 9cm
  =\,\langle \Omega;t_{-\infty}|\{\bar{T}\xi(y_1)\cdots\xi(y_m)\}\{T\xi(x_1)
          \cdots\xi(x_n)\}|\Omega;t_{-\infty}\rangle
\,,\quad
\nonumber
\eea
where $T$ denotes operator time-ordering,
$T\xi(x)\xi(x^\prime)= \theta(x^0\!-\!x^{\prime 0})\xi(x)\xi(x^\prime)
                        + \theta(x^{\prime 0}\!-\!x^0)\xi(x)\xi(x^\prime)$,
and $\bar{T}$ operator anti-time-ordering,
$\bar{T}\xi(x)\xi(x^\prime)= \theta(x^{\prime 0}\!-\!x^0)\xi(x)\xi(x^\prime)
                        + \theta(x^0\!-\!x^{\prime 0})\xi(x)\xi(x^\prime)$.

The exact calculation of $Z[J_+,J_-]$ is impossible for interacting theories but it can be evaluated perturbatively. Writing the action as a free part plus interactions
\beq
S[\xi]=\int d^4x\, \frac{1}{2}\xi\mc{D}_0\xi + S_I[\xi]
\,,
\eeq
we have~\footnote{Formula~(\ref{generating-1}) applies also
for the (fermionic) ghost fields, as can be seen {\it e.g.} from Eq.~(5.1.32)
in Ref.~\cite{Ramond:1981pw}, where $i\Delta$ are the ghost propagators.}
\bea\label{generating-1}
Z[J_-,J_+]&=&{\rm e}^{iS_I\left[\frac{\delta}{i\delta J_+}\right]\,
        - \,iS_I\left[\frac{\delta}{-i\delta J_-}\right] }
  \int[{\cal D}\xi_+{\cal D}\xi_-] \,
  \exp\left\{\int d^4x\Big[\frac{1}{2}
   \left(\begin{smallmatrix}\xi_+,&\xi_-\end{smallmatrix}\right)
\left(\begin{smallmatrix}i\mc{D}_0& 0\\ 0&-i\mc{D}_0\end{smallmatrix}\right)
\left(\begin{smallmatrix}\xi_+\\\xi_-\end{smallmatrix}\right)
+\left(\begin{smallmatrix}\xi_+,&\xi_-\end{smallmatrix}\right)
 \left(\begin{smallmatrix}iJ_+\\-iJ_-\end{smallmatrix}\right)\Big]\right\}
\nonumber\\
&=&{\rm e}^{iS_I\left[\frac{\delta}{i\delta J_+}\right]\,
   - \,iS_I\left[\frac{\delta}{-i\delta J_-}\right] }
\exp\left\{\int d^4xd^4x^\prime\frac{1}{2}
   \left(\begin{smallmatrix}iJ_+,&-iJ_-\end{smallmatrix}\right)(x)
    \left(\begin{smallmatrix}i{\rm \Delta_{++}}&i{\rm \Delta_{+-}}
\\
i{\rm \Delta_{-+}}&i{\rm \Delta_{--}}\end{smallmatrix}\right)(x;x^\prime)
 \left(\begin{smallmatrix}iJ_+\\-iJ_-\end{smallmatrix}\right)(x^\prime)\right\}\,,
\label{generating-2}
\eea
where ${\cal D}_0$ denotes the operator acting on $\xi$
as implied by the free action~(\ref{2nd-w-h}).
The last equality is derived by shifting the fields
\beq\label{shift-xi}
   \left(\begin{smallmatrix}\xi_+\\ \xi_-\end{smallmatrix}\right)
  \rightarrow    \left(\begin{smallmatrix}\xi_+\\ \xi_-\end{smallmatrix}\right)
 - \left(\begin{smallmatrix}i{\rm \Delta_{++}}& i{\rm \Delta_{+-}}\\
                i{\rm \Delta_{-+}}&i{\rm \Delta_{--}}\end{smallmatrix}\right)
   \left(\begin{smallmatrix}iJ_+\\-iJ_-\end{smallmatrix}\right)
\,,
\eeq
where the Keldysh $2\times 2$ matrix of propagators
is defined as the matrix inverse of ${\cal D}_0$,
\beq\label{inverse}
\left(\begin{array}{cc}\mc{D}_0& 0\\ 0&-\mc{D}_0\end{array}\right)(x)
   \left(\begin{array}{cc}i{\rm \Delta_{++}}&i{\rm \Delta_{+-}}
              \\
          i{\rm \Delta_{-+}}&i{\rm \Delta_{--}}\end{array}\right)(x;x^\prime)
    =i\delta^4(x-x^\prime)
\,.
\eeq
The $J$-independent path integrals in~(\ref{generating-2})
contribute phase factors which cancel.
The diagonal elements in (\ref{inverse}) are the (Wightman) Green functions
of the operator $\mc{D}_0$ which appear in usual S-matrix calculations in QFT,
while the off-diagonal elements are solutions of the homogeneous equations,
$\mc{D}_0i\Delta_{-+}=\mc{D}_0i\Delta_{+-}=0$.
These off-diagonal elements in (\ref{inverse}) are necessary
in order to impose the boundary condition $\xi_+=\xi_-$ at time
$t_{out}$ on the $J$-dependent parts of $\xi_+$ and $\xi_-$.

The explicit form of the $\Delta$ propagators can be found by using
(\ref{variational-derivs}) and (\ref{generating-2}), after setting
$S_I\rightarrow 0$. We see that for propagating fields the appropriate elements of
the Keldysh (bosonic) propagators in~(\ref{inverse})
are
\bea
i\Delta_{++}(x;x^\prime\,)
  &=& \langle\Omega| T \xi(x)\xi(x^\prime)|\Omega\rangle
      = \theta(x^0\!-\!x^{\prime 0})i\Delta_{-+}(x;x^\prime)
        + \theta(x^{\prime 0}\!-\!x^0)i\Delta_{+-}(x;x^\prime)
\label{Feynman prop}
\\
i\Delta_{--}(x;x^\prime\,)
  &=& \langle\Omega|\bar T \xi(x)\xi(x^\prime)|\Omega\rangle
      = \theta(x^0\!-\!x^{\prime 0})i\Delta_{+-}(x;x^\prime)
        + \theta(x^{\prime 0}\!-\!x^0)i\Delta_{-+}(x;x^\prime)
\,,
\label{antiFeynman prop}
\eea
where the Wightman functions $i\Delta_{+-}$ and $i\Delta_{-+}$ are defined as,
\beq
i\Delta_{+-}(x;x^\prime)=\langle\Omega|\xi(x^\prime)\xi(x)|\Omega\rangle
\,,\quad
i\Delta_{-+}(x;x^\prime)=\langle\Omega|\xi(x)\xi(x^\prime)|\Omega\rangle
\,,
\label{Wightman functions}
\eeq
and $|\Omega\rangle$ denotes the physical vacuum state
~\footnote{In statistical
field theory pure states are replaced by
more general mixed states (see {\it e.g.}~\cite{Koksma:2009wa,Koksma:2010zi}),
and the evolution of a system is generated by the density operator
$\hat\rho(t)$. In this case the Green functions are defined by the more
general relation,
$i\Delta_{++}(x;x^\prime\,)={\rm Tr}[\hat\rho(t) T \xi(x)\xi(x^\prime)]$,
{\it etc}, where the definitions must be picture independent.
For example, if the density operator $\hat\rho(t)$ is
in the Schr\"odinger picture, the fields $\xi(x)$ must be in the Heisenberg
picture.}. If the fields are non-propagating, the Wightman functions
vanish, $\Delta_{+-}=0=\Delta_{-+}$ and Eq.~(\ref{inverse})
is easily inverted, $i\Delta_{\pm\pm}
           = \pm (\mc{D}_0)^{-1}(x)i\delta^4(x-x^\prime)$.

\subsection{Propagators}

Let us now focus on the problem at hand and in particular on the propagators that correspond to $¨{\cal S}_{\rm F}$ given in (\ref{2nd-w-h}).
The propagators for $w$
\bea
i{\rm \Delta}^{w}_{++}(x;x^\prime)
            &=&\langle\Omega|T w(x)w(x^\prime)|\Omega\rangle
\,,\qquad
i{\rm \Delta}^{w}_{--}(x;x^\prime)
            =\langle\Omega|\bar{T} w(x)w(x^\prime)|\Omega\rangle
\nonumber\\
i{\rm \Delta}^{w}_{-+}(x;x^\prime)
           &=&\langle\Omega|w(x)w(x^\prime)|\Omega\rangle
\,,\qquad\;\;\,
i{\rm \Delta}^{w}_{+-}(x;x^\prime)
            =\langle\Omega|w(x^\prime)w(x)|\Omega\rangle
\,,
\label{w propagators}
\eea
and for $h_{ij}^{TT}$
\bea
i{\rm \Delta}^{ijkl}_{++}(x;x^\prime)
        &=&\langle\Omega| T h_{ij}^{TT}(x)h_{kl}^{TT}(x^\prime)|\Omega\rangle
\,,\qquad
i{\rm \Delta}^{ijkl}_{--}(x;x^\prime)
    =\langle\Omega|\bar{T}h_{ij}^{TT}(x)h_{kl}^{TT}(x^\prime)|\Omega\rangle
\nonumber\\
i{\rm \Delta}^{ijkl}_{-+}(x;x^\prime)
    &=&\langle\Omega|h_{ij}^{TT}(x)h_{kl}^{TT}(x^\prime)|\Omega\rangle
\,,\qquad\;\;\,
i{\rm \Delta}^{ijkl}_{+-}(x;x^\prime)
   = \langle\Omega|h_{ij}^{TT}(x^\prime)h_{kl}^{TT}(x)|\Omega\rangle
\,
\label{hTT propagators}
\eea
can be obtained by solving the equations,
\beq
  \Big(\!-\!\partial_tz^2a^3\frac{\partial_t}{\bar N}
        +\bar N z^2a\nabla^2\Big)\,i{\rm \Delta}^{w}_{\pm\pm}(x;x^\prime)
                          =\pm i\delta^4(x-x^\prime)
\,,\qquad
   \Big(\!-\!\partial_tz^2a^3\frac{\partial_t}{\bar N}
        +\bar N z^2a\nabla^2\Big)\,i{\rm \Delta}^{w}_{\pm\mp}(x;x^\prime)
                          = 0
\,,
\label{w propagators:2}
\eeq
and~\footnote{The normalization on the right hand side can be checked
by noting that, if the graviton were not transverse or traceless,
{\it i.e.} just an ordinary $3\times 3$ symmetric matrix, then
the propagator equation would read
\beq
\frac12\Big[-\partial_ta^3\frac{\partial_t}{\bar N}+\bar N a\nabla^2\Big]
    \,i{\rm \Delta}^{ijkl}_{\pm\pm}(x;x^\prime)
    \rightarrow \pm \frac12(\delta_{ik}\delta_{jl}+\delta_{il}\delta_{jk})
          i\delta^4(x-x^\prime)\nonumber
.
\eeq
Eq.~(\ref{hTT propagators:2}) is obtained from this equation by removing
the longitudinal and the trace degrees of freedom.}
\beq
\Big(\!-\!\partial_ta^3\frac{\partial_t}{\bar N}+\bar N a\nabla^2\Big)
    \,i{\rm \Delta}^{ijkl}_{\pm\pm}(x;x^\prime)
    =\pm (P_{ik}P_{jl}\!+\!P_{il}P_{jk}\!-\!P_{ij}P_{kl})
          i\delta^4(x-x^\prime)
\,,\quad
\Big(\!-\!\partial_ta^3\frac{\partial_t}{\bar N}+\bar N a\nabla^2\Big)
    \,i{\rm \Delta}^{ijkl}_{\pm\mp}(x;x^\prime)
        = 0
\,,
\label{hTT propagators:2}
\eeq
where $z\equiv \sqrt{\epsilon}/3$,
$P_{ij} = \delta_{ij}-\partial_i\partial_j/\nabla^2$ denotes the transverse
projector, and the operator $\nabla^{-2}$ is defined by
$\nabla^{-2}_{\vc x}\delta(\vc x-\vc y) = i\Delta_0(\vc x;\vc y)\equiv
-1/[4\pi\|\vc x-\vc y\|]$.
The projectors on the right hand side of~(\ref{hTT propagators:2})
project out the potentially unphysical (scalar and vector)
`degrees of freedom,'
which are not transverse or which are not traceless, assuring
the correct counting of the physical degrees of freedom.
Due to the infrared problems which plague
Eqs.~(\ref{w propagators:2}--\ref{hTT propagators:2}) in accelerating
space-times,
the corresponding solutions are not simple and need to be handled with
due care~\cite{Janssen:2008px,Janssen:2008dw,Janssen:2009nz}. Here we just note
that, when the fields in (\ref{w propagators}) and~(\ref{hTT propagators})
are Fourier transformed to the spatial momentum space,
\bea
w(x)&=&\int\frac{d^3\vc{k}}{(2\pi)^3}\,e^{-i\vc{k}\cdot\vc{x}}
   \big[w(t,k)a_{\vc{k}}\,+w^*(t,k)a_{-\vc{k}}^+\big]
\nonumber\\
 h^{\rm TT}_{ij}(x)&=&\int\frac{d^3\vc{k}}{(2\pi)^3}\,
  {\rm e}^{-i\vc{k}\cdot\vc{x}}\sum_{\alpha=+,\times}
           \epsilon^\alpha_{ij}(\vc{k})
           \big[h(t,k)a_{\alpha,\vc{k}}+h^*(t,k)a^+_{\alpha,-\vc{k}}
           \big]
\label{spatial Fourier:w+h}
\eea
and written in
conformal time ($\bar N\rightarrow a(\eta), t\rightarrow\eta$),
Eqs.~(\ref{w propagators:2}--\ref{hTT propagators:2}) simplify to,
\bea
 \big(\!-\!\partial_\eta z^2a^2\partial_\eta
     -z^2a^2\vc k^2\big)\,i{\rm \Delta}^{w}_{\pm\pm}(\eta;\eta^\prime;\vc k)
                          &=&\pm i\delta(\eta-\eta^\prime)
\nonumber\\
\big(\!-\!\partial_\eta a^2\partial_\eta - a^2\vc k^2\big)
    \,i{\rm \Delta}^{ijkl}_{\pm\pm}(\eta;\eta^\prime;\vc k)
       &=&\pm (\bar P_{ik}\bar P_{jl}+\bar P_{il}\bar P_{jk}
             -\bar P_{ij}\bar P_{kl})i\delta(\eta-\eta^\prime)
\,,\qquad \bar P_{ij}=\delta_{ij}-\frac{k_ik_j}{\vc k^2}
\,.\quad
\label{w hTT propagators}
\eea
We arrived at these results by choosing the physical state
$|\Omega\rangle$ that is destroyed by the annihilation operators
$a_{\vc{k}}$ and $a_{\alpha,\vc{k}}$ in~(\ref{spatial Fourier:w+h}),
$a_{\vc{k}}|\Omega\rangle =0$, $a_{\alpha,\vc{k}}|\Omega\rangle =0$.
The solutions of Eqs.~(\ref{w hTT propagators})
 can be expressed in terms of the mode functions
$w(\eta,k)$ and $h(\eta,k)$ as,
\bea
 i\Delta^w_{++} &=& \theta(\eta-\eta^\prime) i\Delta^w_{-+}
                + \theta(\eta^\prime-\eta) i\Delta^w_{+-}
\,,\qquad
 i\Delta^w_{+-}=  w^*(\eta,k)w(\eta^\prime,k)
\,,\qquad
 i\Delta^w_{-+}=  w(\eta,k,)w^*(\eta^\prime,k)
\nonumber\\
 i\Delta^{ijmn}_{++} &=& \theta(\eta-\eta^\prime) i\Delta^{ijmn}_{-+}
                + \theta(\eta^\prime-\eta) i\Delta^{ijmn}_{+-}
\nonumber\\
 i\Delta^{ijmn}_{+-}
&=&  \sum_\alpha\epsilon_{ij}^\alpha(\vc{k})\epsilon_{kl}^\alpha(\vc{k})
    h^*(\eta,k)h(\eta^\prime,k)
\,,\quad
 i\Delta^{ijmn}_{-+}
 =\sum_\alpha\epsilon_{ij}^\alpha(\vc{k})\epsilon_{kl}^\alpha(\vc{k})
    h(\eta,k)h^*(\eta^\prime,k)
\,,
\nonumber
\eea
where $w(\eta,k)$ and $h(\eta,k)$ satisfy,
\beq
 \Big(-\partial_\eta^2 - \vc k^2 +\frac{(az)^{\prime\prime}}{az}\Big)
                           (az w(\eta,k)) = 0
\,,\qquad
 \Big(-\partial_\eta^2 - \vc k^2 +\frac{a^{\prime\prime}}{a}\Big)
                           (a h(\eta,k)) = 0
\,,
\label{w hTT mode equation}
\eeq
where $\epsilon_{ij}^\alpha$ denotes the graviton polarization
tensor.~\footnote{The polarization tensor $\epsilon_{ij}^\alpha$ satisfies,
 $\sum_{\alpha=+,\times}\epsilon_{ij}^\alpha\epsilon_{kl}^\alpha
      = (1/2)[\bar P_{ik}\bar P_{jl}+\bar P_{il}\bar P_{jk}
            -\bar P_{ij}\bar P_{kl}]$
and $\sum_{ij}\epsilon_{ij}^\alpha\epsilon_{ij}^\beta
      = \delta^{\alpha\beta}$.}
These equations can be solved for certain choices of the scale factor
$a=a(\eta)$. For example, when $\epsilon=-\dot H/H^2 ={\rm constant}$,
$a=[-(1\!-\!\epsilon)H_0\eta]^{1/(1-\epsilon)}$, where $H_0=H(\eta_0)$,
such that $a^{\prime\prime}/a = (2\!-\!\epsilon)/[(1\!-\!\epsilon)^2\eta^2]$
and~(\ref{w hTT mode equation}) can be solved in terms of the Hankel functions,
\bea
w(k,\eta) &=& \frac{1}{az}\sqrt{\frac{\pi|\eta|}{4}}
   \Big[\alpha_w(k)H^{(1)}_\nu(k|\eta|)+ \beta_w(k)H^{(2)}_\nu(k|\eta|)\Big]
\,,\qquad \nu = \frac{3\!-\!\epsilon}{2(1\!-\!\epsilon)}
\,,\qquad |\alpha_w(k)|^2-|\beta_w(k)|^2=1
\nonumber\\
 h(k,\eta)
    &=& \frac{1}{a}\sqrt{\frac{\pi|\eta|}{2}}
                  \Big[\alpha_h(k)H^{(1)}_\nu(k|\eta|)
                        + \beta_h(k)H^{(2)}_\nu(k|\eta|)\Big]
\,,\qquad |\alpha_h(k)|^2-|\beta_h(k)|^2=1
\,,\qquad
\label{w hTT solution mode equation}
\eea
and where $\nu_w = \nu_h \equiv \nu = (3\!-\!\epsilon)/[2(1\!-\!\epsilon)]$,
and the coefficients $(\alpha_w(k),\beta_w(k))$
and $(\alpha_h(k),\beta_h(k))$ of positive and negative frequency
mode functions are chosen such that the real space propagators
are infrared finite~\cite{Vilenkin:1982wt,Janssen:2009nz}. Alternatively,
one can place the Universe in a large comoving
box~\cite{Tsamis:1993ub,Tsamis:1996qm,Janssen:2008dw,Janssen:2008px},
or in universe with spatially positively curved sections,
which effectively remove the far infrared modes,
thus rendering the state infrared finite. For non-constant $\epsilon$, solutions can be obtained in the slow-roll approximation, by using the de Sitter
mode functions around the time of first Hubble
crossing~\cite{Maldacena:2002vr}, or the methods of \cite{Tsamis:2003zs}.

\medskip

The two-point correlators for all of the auxiliary fields,
$\rho^{ij},\,\rho_\varphi,\, \tilde{n}\,,\tilde{N}^T_i$ and $\tilde{S}$,
can be straightforwardly obtained. It is clear that the off-diagonal elements
(Wightman functions) of~(\ref{inverse}) vanish,
while the diagonal elements are either proportional
to delta functions or involve spatially non-local operators.
In particular, the Green functions for the auxiliary fields as implied
by~(\ref{2nd-w-h}) are~\footnote{Eq.~(\ref{G2}) is obtained by noting
that the inverse of $A_{ijkl}$ is
$(A^{-1})_{ijkl}
 =(1/4)(\delta_{ik}\delta_{jl}+\delta_{il}\delta_{jk}-2\delta_{ij}\delta_{kl})$
in the sense that $A_{ijkl}(A^{-1})_{klmn}
   = \frac12(\delta_{im}\delta_{jn}+\delta_{in}\delta_{jm})$.
That this indeed is the correct definition of the inverse can be checked by
calculating, for example, the propagator for the (scalar) trace
$\rho\equiv\delta_{ij}\rho^{ij}$ of $\rho^{ij}$, for which
the free action~(\ref{2nd-w-h}) yields $\int d^4x[2\bar Na^3H^2\rho^2/3]$.
The corresponding propagator is, of course,
$i\Delta^\rho_{\pm\pm}(x;x^\prime) = \pm[3/(4\bar Na^3H^2)]
 i\delta^4(x-x^\prime)$,
which agrees with the double trace of Eq.~(\ref{G2}), obtained by
applying $\delta_{ij}\delta_{kl}$. One can analogously check
that the propagators for the other components of
$\rho^{ij}$ are correct.}
\bea
i\Delta^{\rho_\varphi}_{\pm\pm}(x;x^\prime)
   &=&\mp \frac{1}{4\bar{N}a^3H^2\epsilon}\,i\delta^4(x-x^\prime)
\label{G1}
\\
i\left(\Delta^\rho_{\pm\pm }\right)_{ijkl}(x;x^\prime)
   &=& \mp \frac{1}{16\bar{N}a^3H^2}
       (\delta_{ik}\delta_{jl}+\delta_{il}\delta_{jk}-2\delta_{ij}\delta_{kl})
                 i\delta^4(x-x^\prime)
\label{G2}
\\
i\Delta^{\tilde n}_{\pm\pm }(x;x^\prime)
   &=&\mp \frac{\bar N}{4(3-\epsilon)a^3H^2}\,i\delta^4(x-x^\prime)
\label{G3}
\\
i\left({\Delta}^{T}_{\pm\pm}\right)_{ij}(x;x^\prime)
   &=&\mp \,a\bar N\,\nabla_x^{-2}P_{ij}
              i\delta^4(x-x^\prime)
\label{G4}
\\
i{\Delta}^{\tilde S}_{\pm\pm}(x;x^\prime)
   &=&\pm \frac{1}{4}a\bar N(3-\epsilon)\nabla_x^{-4}i\delta^4(x-x^\prime)
\label{G5}
\,,
\eea
with all the $i\Delta_{+-}=0$ for the auxiliary fields.
\medskip

Let us now turn to the ghost fields. Their propagators satisfy the equations of motion of the general form,
\beq
\Omega^{\rm F}_{\alpha\beta}
  \left(i\Delta^\eta_{\pm\pm}\right)_{\beta\gamma}(x;x^\prime)
 = \pm i\delta_{\alpha\gamma}\delta^4(x-x^\prime)
\,,\qquad
\Omega^{\rm F}_{\alpha\beta}
  \left(i\Delta^\eta_{\pm\mp}\right)_{\beta\gamma}(x;x^\prime)
 = 0
\,.
\eeq
Whether the ghosts are propagating or not depends of course on the choice of gauge conditions and in particular whether the latter contain time derivatives. In the former case we have
\bea
(i\Delta^\eta_{\alpha\beta})_{++}(x;x^\prime)
 &=& \langle\Omega|T{\eta}_\alpha(x)\bar\eta_\beta(x^\prime)|\Omega\rangle
  = \theta(x^0-x^{\prime 0})
          (i\Delta^\eta_{\alpha\beta})_{-+}(x;x^\prime)
    + \theta(x^{\prime 0}-x^0)
          (i\Delta^\eta_{\alpha\beta})_{+-}(x;x^\prime)
\nonumber\\
(i\Delta^\eta_{\alpha\beta})_{--}(x;x^\prime)
 &=&  \langle\Omega|\bar T\eta_\alpha(x)\bar\eta_\beta(x^\prime)|\Omega\rangle
  = \theta(x^0-x^{\prime 0})
          (i\Delta^\eta_{\alpha\beta})_{+-}(x;x^\prime)
    + \theta(x^{\prime 0}-x^0)
          (i\Delta^\eta_{\alpha\beta})_{-+}(x;x^\prime)
\,,
\label{fermionic Feynman}
\eea
where
\beq
(i\Delta^\eta_{\alpha\beta})_{+-}(x;x^\prime)
  =  - \,\langle\Omega|\bar\eta_\beta(x^\prime)\eta_\alpha(x)|\Omega\rangle
\,,\qquad
(i\Delta^\eta_{\alpha\beta})_{-+}(x;x^\prime)
  =  \langle\Omega|\eta_\alpha(x)\bar\eta_\beta(x^\prime)|\Omega\rangle
\,.
\label{fermionic Wightman}
\eeq
Note that because they are anticommuting, their corresponding time ordered and anti-time
ordered propagators differ slightly from the
bosonic counterparts (\ref{Feynman prop}--\ref{antiFeynman prop}).
Since common gauge choices do not lead to propagating
ghosts, the corresponding propagators have simpler form than the one given
in~(\ref{fermionic Feynman}--\ref{fermionic Wightman}).
Consider, for example, the {\it tensor gauge}
\beq
\mc{Q}_0=h\,,\quad \mc{Q}_i
  = \partial_j\left(h_{ij}-\frac{\delta_{ij}}{3}h\right)
\,.
\label{gauge-cond1}
\eeq
Imposing the gauge conditions $\mc{Q}_\alpha=0$ sets $h=\tilde{h}=h^T_i=0$
and hence the metric perturbation contains only the (transverse traceless)
tensor part of the spatial metric, while $w$ represents the perturbation
of the inflaton. In linear theory it corresponds to the uniform spatial
expansion gauge, or the comoving gauge.

 From Eq.~(\ref{ghosts:tensor gauge}) in Appendix~C it follows that
the free part of the ghost operator $\Omega^F_{\alpha\beta}$
is non-dynamical in the tensor gauge; it reads:
\beq
\Omega^{\rm F}=\left(\begin{array}{cc}
                       -6H & -2a^{-2}\partial_j^x \\
                  0 & -a^{-2}\left(\delta_{ij}\nabla_x^2
                          +\frac{1}{3}\partial^x_i\partial^x_j\right)
                       \end{array}\right)\delta(\vc x-\vc y)
\,.
\label{OmegaF:tensor gauge}
\eeq
The ghost equations, when written in a condensed matrix notation:
 $\Omega^{\rm F}\cdot (i\Delta^\eta)_{\pm\pm} = \pm i\delta^4$,
$\Omega^{\rm F}\cdot (i\Delta^\eta)_{\pm\mp}=0$,
imply,
\beq
 (i\Delta^\eta)_{\pm\pm}(x;x^\prime)
      = \left(\begin{array}{cc}
        (i\Delta_{00}^\eta)_{\pm\pm}(x;x^\prime)  &
         (i\Delta_{0j}^\eta)_{\pm\pm}(x;x^\prime)   \\
         (i\Delta_{i0}^\eta)_{\pm\pm}(x;x^\prime) &
         (i\Delta_{ij}^\eta)_{\pm\pm}(x;x^\prime)
                       \end{array}\right)
      = \left(\begin{array}{cc}
           \mp\frac{1}{6H} &
           \pm\frac{1}{4H}\frac{\partial_j^x}{\nabla^2_x} \\
           0 &
           \mp a^2\left(\delta_{ij}
            -\frac{1}{4}\frac{\partial^x_i\partial^x_j}{\nabla_x^2}\right)
                   \frac{1}{\nabla_x^2}
       \end{array}\right)i\delta^4(x-x^\prime)
\,
\label{ghost propagator:tensor gauge}
\eeq
and $(i\Delta^\eta)_{\pm\mp}(x;x^\prime)=0$.

The second gauge we consider here is the {\it uniform field gauge},
\beq
\mc{Q}_0=\varphi\,,\quad \mc{Q}_i
  = \partial_j\left(h_{ij}-\frac{\delta_{ij}}{3}h\right)
\,.
\label{gauge-cond2}
\eeq
In this gauge $\mc{Q}_\alpha=0$ fixes $\varphi=\tilde{h}=h^T_i=0$
and hence the metric perturbation contains the tensor
and the spatial trace perturbation. This means
that $w$ now represents the perturbation
of the local spatial volume, or the spatial curvature perturbation. From Eq.~(\ref{ghosts:uniform field gauge}) we see
that, also in the uniform field gauge, the free ghost operator is nondynamical,
\beq
\Omega^{\rm F}=\left(\begin{array}{cc}
                       -2\sqrt{\epsilon}H & 0 \\
                  0 & -a^{-2}\left(\delta_{ij}\nabla_x^2
                          +\frac{1}{3}\partial^x_i\partial^x_j\right)
                       \end{array}\right)\delta(\vc x-\vc y)
\,,
\label{OmegaF:ufg}
\eeq
implying the following 2-point ghost correlators
\beq
 (i\Delta^\eta)_{\pm\pm}(x;x^\prime)
      = \left(\begin{array}{cc}
           \mp \frac{1}{2\sqrt{\epsilon}H} &
           0 \\
           0 &
           \mp a^2\left(\delta_{ij}
                 -\frac{1}{4}\frac{\partial^x_i\partial^x_j}{\nabla_x^2}\right)
                    \frac{1}{\nabla_x^2}
             \end{array}\right)i\delta^4(x-x^\prime)
\,;\qquad
 (i\Delta^\eta)_{\pm\mp}(x;x^\prime)=0
\,.
\label{ghost propagator:ufg}
\eeq
Notice that the gauge~(\ref{gauge-cond2}) is not suitable when one is
interested
in the de Sitter limit $\epsilon\rightarrow 0$, since the ghost
propagator~(\ref{ghost propagator:ufg}) appears singular in that limit.
That problem can be fixed by simply replacing $Q_0$ in~(\ref{gauge-cond2})
by $Q_0=\varphi/\dot\phi$. In order to make the whole
action~(\ref{action pert}--\ref{S_I}) regular, one also needs
to redefine the canonical
momentum of $\varphi$ as ${\cal P}_\phi\pi_\varphi\rightarrow \pi_\varphi$,
and similarly $-(\sqrt{\epsilon}/3)w\rightarrow\tilde\varphi$,
such that the corresponding propagators~(\ref{G1}) and~(\ref{w propagators:2})
become regular. When these changes are exacted,
the de Sitter limit of our
path integral~(\ref{transition Amplitude3}) will be regular
in the uniform field gauge.

\vskip 0.1cm

\subsection{Diagrammatic rules}

After expanding (\ref{generating-2}) in powers of $S_I$, we are led to
a diagrammatic expansion for $Z[J_+,J_-]$, analogous to that of the
vacuum-to-vacuum amplitude in the presence of a source $J$ of standard QFT.
Diagrams are now composed of external lines coupling $+$ and $-$ currents
at external points to $+$ and $-$ vertices, as well as internal lines coupling
$+$ and $-$ vertices with each other. Each vertex carries a $+i$ or $-i$
factor respectively, along with the relevant interaction factors.
Vertices and currents of the same valence,
($+,+$ and $-,-$) are joined by lines corresponding to
$\Delta_{++}$ and $\Delta_{--}$ propagators, respectively,
while those of opposite valence ($+,-$ and $-,+$) are joined by
$\Delta_{+-}$ and $\Delta_{-+}$ propagators, respectively.
Finally, one integrates over the coordinates of internal points.

It is important to keep in mind that, as mentioned above,
the 2-point correlators of the auxiliary fields and of the ghosts
in common gauges satisfy
${\rm \Delta}_{+-}={\rm \Delta}_{-+}=0$ while the
$\Delta_{++}$ and $\Delta_{--}$ are either delta functions or involve
non-local spatial operators of the type
$\delta(t-t^\prime)/\|\vc x-\vc x^\prime\|$ or even
$\delta(t-t^\prime)\nabla^{-2}\|\vc x-\vc x^\prime\|^{-1}
=-(4\pi)^{-1}\delta(t-t^\prime)\int d^3\vc z \|\vc x-\vc z\|^{-1}
           \|\vc z-\vc x^\prime\|^{-1}$ as in Eqs.~(\ref{G4}--\ref{G5}).
Since there is no actual propagation in spacetime,
these correlators always correspond to effective vertices,
either local or non-local.

Once $Z[J_-,J_+]$ has been expressed as a sum of diagrams to the desired order in the interaction or in loops, variational derivatives will then give expectation values according to (\ref{variational-derivs}). Note that in case all fields in (\ref{variational-derivs}) are taken at the same time, the valence of the currents is immaterial; here we will be using $J_+$, so that {\it e.g.}
\beq
\langle\Omega|\xi(t,\vc{x})\xi(t,\vc{y})|\Omega\rangle
 = \frac{\delta}{i\delta J_+(t,\vc{x})}\frac{\delta}{i\delta J_+(t,\vc{y})}
   Z[J_-,J_+]\Big|_{J_-=J_+=0}
\,,
\label{2-point}\eeq
and
\beq
\langle\Omega|\xi(t,\vc{x})\xi(t,\vc{y})\xi(t,\vc{z})|\Omega\rangle
  = \frac{\delta}{i\delta J_+(t,\vc{x})}\frac{\delta}{i\delta J_+(t,\vc{y})}
  \frac{\delta}{i\delta J_+(t,\vc{z})} Z[J_-,J_+]\Big|_{J_-=J_+=0}
\,.
\label{3-point}\eeq

 All of the above mirrors the corresponding discussion for standard QFT
transition amplitudes. We can thus summarize the diagrammatic rules for the
calculation of an $n$-point function
$\langle\Omega|\xi(t,\vc{x})\xi(t,\vc{y})\ldots\xi(t,\vc{z})|\Omega\rangle$
as follows (see the appendix of \cite{Weinberg:2005vy}
for a complete definition of such rules):
\begin{itemize}
\item{draw the usual Feynman Diagrams with all external points
  having a + valence;}
\item{all internal vertices are either + or - and carry a $+i$ or $-i$
factor, respectively, along with all the relevant interaction operators;}
\item{connect external points with vertices and vertices with each other by
      the propagators of the appropriate valence;}
\item{integrate over the temporal and spatial coordinates of all the vertices.}
\end{itemize}

\subsection{The tensor gauge and cosmological perturbations}

The N-point functions of the trace of the spatial metric
$\langle\Omega| h(t,\vc{x}_1)h(t,\vc{x}_2)\ldots h(t,\vc{x}_N)|\Omega \rangle$
on comoving spacelike hypersurfaces are significant cosmological observables, particularly in relation with the non-Gaussianity of cosmological perturbations. During inflation, such quantities can be directly computed in the uniform field gauge, since in that gauge $w\rightarrow h$. However, the interaction terms in ${\cal S}_{\rm I}$ are simplest in the tensor gauge and the question arises of how to relate results in these two different gauges. The issue is that, although the transition amplitude (\ref{transition Amplitude3}) is independent of the choice of gauge conditions, simply coupling $w$ to a current and calculating expectation values does not automatically lead to a gauge invariant expression. This is obvious since $w$ is not invariant under non-linear transformations and adding a $Jw$ term in the action explicitly breaks its gauge invariance. A more physical way to understand this is to note that a time slice defined by some fixed value $t$ of the background time corresponds to \emph{physically different} hypersurfaces in different gauges. Thus, one cannot simply compare $\langle w^N(t) \rangle$ for fixed $t$ in different gauges because it refers to physically different quantities evaluated on different timelike hypersurfaces. Such difficulties can be circumvented however if we define the observables of interest and the timelike hypersurfaces on which they are to be calculated in a geometrical
fashion~\cite{Gasperini:2009wp}.

So, let us choose the tensor gauge to obtain the interaction terms. The physical interpretation of $w$ is that it represents the quanta of the scalar field: $w=-({3}/{\sqrt{\epsilon}})\varphi$, while the metric perturbations are transverse traceless tensors. Interactions between these degrees of freedom are specified by ${\cal S}_{\rm I}$ in this gauge, given explicitly in section \ref{The 3-point and 4-point functions of Inflaton perturbations} below. The transition amplitude that enters in (\ref{in-in}) is then written as
\beq\label{transition Amplitude3b}
\langle\Omega; t_{+\infty}|\Omega;t_{-\infty}\rangle \,
= \int 
\,\,[{\cal D}{w} {\cal D}h^{TT}_{ij}{\cal D}\sigma {\cal D}
\bar{\eta}{\cal D}\eta]\,
\, {\rm e}^{\,{\rm i}{\cal S}_{\rm F}\,+\,{\rm i}\!
 \int{\rm d}^4\!x \,\bar{\eta}_\alpha{\Omega}^{\rm F}_{\alpha\beta}\eta_\beta\,
+\,{\rm i}{\cal S}_{\rm I}\,
+\,{\rm i}\!\int{\rm d}^4\!x \,
       \bar{\eta}_\alpha{\Omega}^{\rm I}_{\alpha\beta}\eta_\beta}
\,,
\eeq
where the operator $\Omega_{\alpha\beta}$ is the one appropriate for the particular gauge choice -- see (\ref{OmegaF:tensor gauge}) and (\ref{ghosts:tensor gauge}) -- and the vacuum states have been obtained by considering large time intervals and the appropriate $i{\rm \epsilon}$ prescription. The expectation value of any product of $w$'s can now be calculated using the diagrammatic rules described above, remembering that the meaning of correlators in this gauge is
\beq
\langle\Omega|w(t,\vc{x}_1)w(t,\vc{x}_2)\ldots w(t,\vc{x}_N)|\Omega\rangle
 = \left(-\frac{3}{\sqrt{\epsilon}}\right)^N
 \langle\Omega| \varphi(t,\vc{x})\varphi(t,\vc{x}_2)\ldots
       \varphi(t,\vc{x}_N)|\Omega\rangle\,.
\eeq

Let us now consider the hypersurfaces defined by the condition $\varphi = 0$,
which are of course different from the $t=\rm constant$ hypersurfaces in the tensor gauge we are using. We are interested in the determinant ${\rm Det}[\gamma_{ij}]_{\varphi=0}$ of the 3-metric $\gamma_{ij}$ induced on these hypersurfaces which measures their local expansion. Writing
\beq
{\rm Det}(\gamma_{ij})_{\varphi=0}=a(t)^2e^{2\zeta}
\eeq
we can calculate the expectation value of any N-point function of $\zeta$ from
\beq\label{gamma-Npoint-correlator}
\langle\Omega| \zeta(t,\vc{x}_1)\zeta(t,\vc{x}_2)\ldots\zeta(t,\vc{x}_N)|\Omega \rangle
= \zeta\left(\frac{\delta}{i\delta J_+(t,\vc{x}_1)}\right)
 \zeta\left(\frac{\delta}{i\delta J_+(t,\vc{x}_2)}\right)\ldots
\zeta\left(\frac{\delta}{i\delta J_+(t,\vc{x}_N)}\right)
      Z[J_-,J_+]\Big|_{J_-=J_+=0}\,,
\eeq
where on the {r.h.s.} $\zeta(w)$
has been expressed as the appropriate function of $w$
and all occurrences of $w$
have been replaced by $\delta/[i\delta J_+(t,\vc{x})]$. For example, to second
order in perturbations, $\zeta$ is related to $w$ in the tensor gauge by
 ({\it cf.} {\it e.g.}~\cite{Maldacena:2002vr}):
\bea
36 \,\zeta &\simeq& 6w + \left(\epsilon-\frac{1}{2}\eta\right)w^2
\nonumber\\
&& + \frac{1}{H} \dot{w}w - \frac{1}{4}\frac{1}{a^2H^2}
\left((\partial_iw)(\partial_iw)
 -\frac{\partial_i\partial_j}{\nabla^2}\Big((\partial_iw)(\partial_jw)\Big)
\right)
+\frac{\epsilon}{H}(\partial_iw)\Big(\frac{\partial_i}{\nabla^2}\dot{w}\Big)
 -\frac{\epsilon}{H}\frac{\partial_i\partial_j}{\nabla^2}
 \left((\partial_iw)\Big(\frac{\partial_j}{\nabla^2}\dot{w}\Big)\right)
\nonumber\\
&&-\frac{3}{2H}\dot{h}^{\rm TT}_{ij}\partial_i\partial_jw\,.
\label{gamma}
\eea
When restricted to long wavelengths, formula (\ref{gamma-Npoint-correlator}) could be considered as a quantum mechanical generalization of the $\Delta N$ formula. Analogous expressions can be used for relating any quantity of interest to results obtained in the tensor gauge.

\section{The 3-point and 4-point functions of Inflaton perturbations}
\label{The 3-point and 4-point functions of Inflaton perturbations}

As an illustration of the formalism developed above, we will derive expressions
for the tree-level 3-point and 4-point functions of the inflaton
fluctuations.
We shall work in the tensor gauge~(\ref{gauge-cond1}). 
All of the relevant cubic vertices are presented in
Appendix~D in Eqs.~(\ref{H:cubic}--\ref{L_0:tensor gauge}). Here
we list the specific couplings that are relevant for computation
of three and four point functions.

From Eqs.~(\ref{H:cubic}--\ref{L_0:tensor gauge}) we can easily extract
the terms involving $w^3$:~\footnote{The result~(\ref{cubic-h=0}) can be compared to Eq.~(3.8) of
Ref.~\cite{Maldacena:2002vr}, which, when expressed in our
language, $\dot\phi\rightarrow 2\sqrt{\epsilon}\,H$,
$\dot\rho\rightarrow H$, $\varphi\rightarrow
-(\sqrt{\epsilon}/3)w$, $\dot\varphi\rightarrow
(-\sqrt{\epsilon}/3)(\dot w+(\epsilon-\eta)Hw)$,
$\nabla^2\chi\rightarrow (\epsilon/3)\dot w$, reduces to,
\bea
 {\cal S}_{www}|_{\rm Maldacena}\!\! &=& \!\int d^3x dt
 \,\frac{a^3}{9}\bigg\{
     \Big(\frac{\epsilon^2}{2}-\frac{\epsilon^3}{6}\Big)w\dot w^2
     + \frac{\epsilon^2}{3}w(\partial_i \dot w)\Big(\frac{\partial_i}{\nabla^2}\dot w\Big)
     + \frac{\epsilon^3}{6}w\Big(\frac{\partial_i\partial_j}{\nabla^2} \dot w\Big)
          \Big(\frac{\partial_i\partial_j}{\nabla^2} \dot w\Big)
\label{Swww:Maldacena}\nonumber\\
 && \hskip 1.cm
   +\,\Big(\frac{\epsilon^3}{6}-\frac{\epsilon^2\eta}{2}\Big)Hw^2\dot w
    +\Big(-\epsilon^3+\frac{\epsilon^2V_{,\phi\phi}}{6H^2}
    +\frac{\epsilon^{3/2}V_{,\phi\phi\phi}}{18H^2}
    +\frac{\epsilon^2(\epsilon+\eta)^2}{6}\Big)H^2w^3
    +\frac{\epsilon^2}{6}w\Big(\frac{\nabla w}{a}\Big)^2
 \bigg\}
. \nonumber
\eea
 We extracted $1/9$ in~(\ref{Swww:Maldacena}) as opposed to $1/18$ in~(\ref{cubic-h=0}) in order to
 compensate for the different definition of the Planck constant in Ref.~\cite{Maldacena:2002vr},
 where $8\pi G_N=1$ (recall that here $16\pi G_N=\kappa^2=1$). By
 comparing Eqs.~(\ref{cubic-h=0}) and~(\ref{Swww:Maldacena}) we see that the leading
 order (${\cal O}(\epsilon^2)$) terms agree, while there are some
 disagreements in the subleading terms
(recall that $V_{,\phi\phi}$ and
$V_{,\phi\phi\phi}/\sqrt{\epsilon}$ are both linear in slow roll
parameters). Since it involves subdominant terms,
this discrepancy is quantitatively unimportant.
Nevertheless, it would be of interest to find the source of the disagreement.}
\bea S_{www}\!\! &=& \!\!\int {\rm d}^3\!x{\rm d}t
\,\frac{\bar{N}a^3}{18} \Bigg\{
\left(\frac{\epsilon^2}{2}-\frac{\epsilon^3}{12}\right)w\dot{w}^2
      +\frac{\epsilon^2}{3}w(\partial_i\dot{w})\left(\frac{\partial_i}{\nabla^2}\dot{w}\right)
 +\frac{\epsilon^3}{12}\,w\left(\frac{\partial_i\partial_j}{\nabla^2}
  \dot{w}\right)\left(\frac{\partial_i\partial_j}{\nabla^2}\dot{w}\right)
\label{cubic-h=0}\\
&&\hspace{2.cm}+\,\left(\frac{\epsilon^3}{3}-\frac{\epsilon^2\eta}{2}\right)
     H w^2\dot{w}
+\left(-\frac{\epsilon^3}{2}+\frac{\epsilon^2V_{,\phi\phi}}{6H^2}
+\frac{\epsilon^{3/2}V_{,\phi\phi\phi}}{9H^2}
 +\frac{\epsilon^2\eta^2}{6}\right)H^2 w^3
     +\frac{\epsilon^2}{6}w\Big(\frac{\nabla w}{a}\Big)^2
 \Bigg\}
\,. \nonumber \eea
Notice that the $w^2\dot w$  term can be integrated by
parts to obtain terms that can be combined with the $w^3$ terms.
The terms in Eqs.~(\ref{H:cubic}--\ref{L_0:tensor gauge}) of relevance for
the tensor three-point function involving three gravitons
are,
\bea
 {\cal S}_{hhh}\!\! &=& \!\int d^3x dt
 \,\bar Na^3\bigg\{
      - \frac12h_{ij}^{TT}\dot h_{jl}^{TT} \dot h_{li}^{TT}
      - 3Hh_{ij}^{TT}h_{jl}^{TT} \dot h_{li}^{TT}
      - \frac{2(5-\epsilon)H^2}{3}h_{ij}^{TT}h_{jl}^{TT} h_{li}^{TT}
\label{Shhh}\\
 && \hskip 1.cm
   +\,\frac{h_{ij}^{TT}}{a^2}
   \Big(\frac{1}{4}(\partial_ih_{kl}^{TT})(\partial_jh_{kl}^{TT})
   + \frac{1}{2}(\partial_lh_{jk}^{TT})(\partial_kh_{il}^{TT})
   - \frac{3}{2}(\partial_lh_{ik}^{TT})(\partial_lh_{jk}^{TT})\Big)
 \bigg\}
 \,.
\nonumber
 \eea
There are also the cubic terms that involve two scalars and one
graviton
and those that involve one scalar and two gravitons,
%
\bea
 {\cal S}_{wwh}\!\! &=& \!\int d^3x dt
 \,\bar Na^3\bigg\{
     - \frac{\epsilon^2}{36}(\dot w h_{ij}^{TT}+ w\dot h_{ij}^{TT})
                     \frac{\partial_i\partial_j}{\nabla^2}\dot w
     + \frac{\epsilon^2}{36}h_{ij}^{TT}\Big(\frac{\partial_j\partial_l}{\nabla^2}\dot w\Big)
                     \Big(\frac{\partial_l\partial_i}{\nabla^2}\dot w\Big)
      + \frac{\epsilon}{18}h_{ij}^{TT}\Big(\frac{\partial_iw}{a}\Big)
                     \Big(\frac{\partial_jw}{a}\Big)
\bigg\}
\label{Swwh}\\
 {\cal S}_{whh}\!\! &=& \!\int d^3x dt
 \,\bar Na^3\bigg\{
     \frac{\epsilon}{24}w \dot  h_{ij}^{TT}\dot h_{ij}^{TT}
      -\frac{\epsilon}{6}\dot w  h_{ij}^{TT}\dot h_{ij}^{TT}
      +\frac{\epsilon}{6}\Big(\frac{\partial_i\partial_j}{\nabla^2}\dot w\Big)  h_{jl}^{TT}\dot h_{li}^{TT}
      - \frac{\epsilon}{12}\Big(\frac{\partial_i}{\nabla^2}\dot w\Big)(\partial_i h_{jl}^{TT})\dot h_{jl}^{TT}
\nonumber\\
  && +\,\frac{\epsilon H}{3}w \dot  h_{ij}^{TT}h_{ij}^{TT}
      -\frac{\epsilon H}{6}\dot w  h_{ij}^{TT}h_{ij}^{TT}
      +\frac{\epsilon H}{3}\Big(\frac{\partial_i\partial_j}{\nabla^2}\dot w\Big)  h_{jl}^{TT}h_{li}^{TT}
\nonumber\\
   && +\, \frac{\epsilon(7+\epsilon-2\eta)\epsilon H^2}{6} w h_{ij}^{TT}h_{ij}^{TT}
   - \frac{\epsilon}{6}wh_{ij}^{TT}\Big(\frac{\nabla^2}{a^2}h_{ij}^{TT}\Big)
   - \frac{\epsilon}{8}w\Big(\frac{\partial_lh_{ij}^{TT}}{a}\Big)
             \Big(\frac{\partial_lh_{ij}^{TT}}{a}\Big)
 \bigg\}
 \,.
\label{Swhh}
\eea
Notice that the last term in~(\ref{Swwh}) will contribute
as ${\cal O}(\epsilon^2)$ to the 4-point function~\cite{Seery:2008ax}.
For the calculation of the scalar four-point functions
the terms involving $w^2$ and any of the auxiliary fields
$\{\tilde S,\tilde n,\delta N_i^T,\rho^{ij},\rho_\varphi\}$ are
also needed,
\bea
 S_{ww(aux)} &=& \int {\rm d}^3\!x{\rm d}t \bar Na^3
\Bigg\{\frac{\tilde n}{\bar N}
 \bigg[
      \Big(-\frac{\epsilon}{18}+\frac{\epsilon^2}{36}\Big)\dot w^2
      -\frac{\epsilon(\epsilon-\eta)}{9}Hw\dot w
      +\Big(\frac{\epsilon^2}{2}-\frac{\epsilon^2\eta}{9}-\frac{\epsilon\eta^2}{18}\Big)H^2w^2
      - \frac{\epsilon V_{,\phi\phi}}{18}w^2
\nonumber\\
  &&\hskip 2.cm  -\,\frac{\epsilon}{18}\Big(\frac{\partial_iw}{a}\Big)
                                     \Big(\frac{\partial_iw}{a}\Big)
            -\, \frac{\epsilon^2}{36}\Big(\frac{\partial_i\partial_j}{\nabla^2}\dot w\Big)
                                     \Big(\frac{\partial_i\partial_j}{\nabla^2}\dot w\Big)
            - \frac{\epsilon^2H}{9}(\partial_iw)\Big(\frac{\partial_i}{\nabla^2}\dot w\Big)
 \bigg]
\nonumber\\
&+&\!\!\frac{\nabla^2\tilde S}{(3-\epsilon)\bar Na^2H}
 \bigg[
      \Big(\frac{\epsilon}{18}-\frac{\epsilon^2}{36}\Big)\dot w^2
      +\Big(\frac{(1-\epsilon)\epsilon^2}{18}-\frac{\epsilon\eta}{9}\Big)Hw\dot w
      +\Big(-\frac{\epsilon^2}{2}-\frac{\epsilon\eta}{6}+\frac{2\epsilon^3}{9}
            +\frac{\epsilon^2\eta}{6}+\frac{\epsilon\eta^2}{18}\Big)H^2w^2
\nonumber\\
  &&\hskip 1.9cm +\,\frac{\epsilon V_{,\phi\phi}}{18}w^2
            + \frac{\epsilon}{18}\Big(\frac{\partial_iw}{a}\Big)
                                     \Big(\frac{\partial_iw}{a}\Big)
            + \frac{\epsilon^2}{36}\Big(\frac{\partial_i\partial_j}{\nabla^2}\dot w\Big)
                                     \Big(\frac{\partial_i\partial_j}{\nabla^2}\dot w\Big)
            + \frac{\epsilon^2H}{9}(\partial_iw)\Big(\frac{\partial_i}{\nabla^2}\dot w\Big)
 \bigg]
\nonumber\\
  && +\, \frac{\epsilon^2H}{9}w\rho^{ij}\Big(\frac{\partial_i\partial_j}{\nabla^2}\dot w\Big)
     - \frac{\epsilon^2H^2}{9}\rho w^2
     + \rho_\varphi
            \bigg[
               - \frac{\epsilon^2H}{9}w\dot w
               + \frac{\epsilon^2\eta H^2}{9}w^2
               + \frac{\epsilon^2H}{9}(\partial_iw)\Big(\frac{\partial_i}{\nabla^2}\dot w\Big)
           \bigg]
\nonumber\\
   &&  +\,
    \frac{\epsilon^2(\partial_{(i}\tilde N^T_{j)}+\partial_i\partial_j\tilde S)}{18\bar Na^2}w
                                 \Big(\frac{\partial_i\partial_j}{\nabla^2}\dot w\Big)
     - \frac{\epsilon(\tilde N^T_{i}+\partial_i\tilde S)}{9\bar Na^2}\dot w(\partial_iw)
\bigg\}
\,.
\label{cubic-aux}
\eea
The purely quartic scalar terms can be read off from
Eqs.~(\ref{nC0:quartic}--\ref{quartic hamiltonian}) in Appendix~D,
\beq S_{wwww} = \int {\rm d}^3\!x{\rm d}t \bar Na^3\,
\bigg\{-\frac{\epsilon^2V_{,\phi\phi\phi\phi}}{1944}w^4
            - \frac{\epsilon^{5/2}V_{,\phi\phi\phi}}{972}w^4
\bigg\}
\,.
\label{Swwww}
\eeq
With the above interaction vertices one can, for example, compute
the scalar contributions to the 3-point and 4-point functions, as
we describe below.

\subsection{The 3-point function $\langle w^3\rangle$}
\label{The 3-point function}

 We can now use the diagrammatic rules described above
to obtain the tree level 3-point function. The relevant diagrams
are
\beq\parbox{20mm}{
\begin{fmffile}{cubic1+}
\begin{fmfgraph*}(30,30)
\fmfleftn{i}{2}\fmfrightn{o}{1}
\fmflabel{+}{i1}
\fmflabel{+}{i2}
\fmflabel{+}{o1}
\fmf{plain}{i1,v1,i2}
\fmf{plain}{v1,o1}
\fmflabel{+}{v1}
\end{fmfgraph*}
\end{fmffile}}
+\qquad\parbox{20mm}{
\begin{fmffile}{cubic1-}
\begin{fmfgraph*}(30,30)
\fmfleftn{i}{2}\fmfrightn{o}{1}
\fmflabel{+}{i1}
\fmflabel{+}{i2}
\fmflabel{+}{o1}
\fmf{plain}{i1,v1,i2}
\fmf{plain}{v1,o1}
\fmflabel{-}{v1}
\end{fmfgraph*}
\end{fmffile}}
\eeq
where the vertex is determined by (\ref{cubic-h=0}). Keeping the
leading order (${\cal O}(\epsilon^2)$) slow-roll terms in
Eq.~(\ref{cubic-h=0}),
\beq
 S_{www} = \int {\rm d}^3\!x{\rm d}t
\,\bar{N}a^3\bigg\{
             \frac{\epsilon^2}{36}w\dot{w}^2
             +\frac{\epsilon^2}{54}w(\partial_i\dot{w})\left(\frac{\partial_i}{\nabla^2}\dot{w}\right)
             + \frac{\epsilon^2}{108}w\Big(\frac{\partial_i w}{a}\Big)\Big(\frac{\partial_i w}{a}\Big)
\bigg\}
\,.
\label{cubic-SR}
\eeq
and recalling Eqs.~(\ref{transition Amplitude2}),
(\ref{in-in}--\ref{variational-derivs}) and~(\ref{3-point}) we see
that the first interaction term in~(\ref{cubic-SR}) contributes as
\bea
  \langle\Omega| w(t,\vc{x})w(t,\vc{y})w(t,\vc{z})|\Omega\rangle^{(1)}
   &=& i\int\limits_{-\infty}^{+\infty}\!\!d^3\vc{u}\,\bar{N}d\tau \,
a^3\frac{\epsilon^2}{36} \bigg\{\langle\Omega| T
w(t,\vc{x})w(\tau,\vc{u})|\Omega\rangle
\frac{\partial}{\bar{N}\partial\tau} \langle\Omega| T
w(t,\vc{y})w(\tau,\vc{u})|\Omega\rangle
 \nonumber\\
 &&\hskip 2.5cm
 \times\,\frac{\partial}{\bar{N}\partial \tau} \langle\Omega| T w(t,\vc{z})w(\tau,\vc{u})|\Omega\rangle
 + \text{perms}
  \nonumber\\
&&\hskip -1cm
-\langle\Omega|
w(\tau,\vc{u})w(t,\vc{x})|\Omega\rangle
  \frac{\partial}{\bar N\partial\tau}\langle\Omega| w(\tau,\vc{u})w(t,\vc{y})|\Omega\rangle
 \frac{\partial}{\bar N\partial \tau} \langle\Omega| w(\tau,\vc{u})w(t,\vc{z})|\Omega\rangle
 + \text{perms}
 \bigg\}
 \nonumber\\
&=&\!\!\int\limits_{-\infty}^{t}\!\!d^3\vc{u}\,d\tau \,\, a^3
\frac{\epsilon^2}{36\bar N}\bigg\{
\Delta^w_{+-}(\tau,\vc{u};t,\vc{x})\partial_\tau\Delta^w_{+-}(\tau,\vc{u};t,\vc{y})
\partial_\tau\Delta^w_{+-}(\tau,\vc{u};t,\vc{z})
 + \text{perms}
\nonumber\\
&&-\Delta^w_{-+}(\tau,\vc{u};t,\vc{x})
\partial_\tau\Delta^w_{-+}(\tau,\vc{u};t,\vc{y})
\partial_\tau\Delta^w_{-+}(\tau,\vc{u};t,\vc{z})
  + \text{perms}
\bigg\}
\,,
\label{3point:1}
 \eea
which is just the retarded contribution. Including in the similar
manner the other two leading order contributions
from~(\ref{cubic-SR}) we finally get for the scalar three point
function,
\bea
\langle\Omega|w(t,\vc{x})w(t,\vc{y})w(t,\vc{z})|\Omega\rangle
   &=& \int\limits_{-\infty}^{t}\!\!d^3\vc{u}\,d\tau \,
   \frac{a^3\epsilon^2}{18\bar N}
   \bigg\{
\Delta^w_{+-}(\tau,\vc{u};t,\vc{x})
  \bigg[\frac12\partial_\tau\Delta^w_{+-}(\tau,\vc{u};t,\vc{y})
        \partial_\tau\Delta^w_{+-}(\tau,\vc{u};t,\vc{z})
\nonumber\\
&&\hskip -3.5cm
      +\,\frac13\frac{\partial}{\partial u^i}\partial_\tau\Delta^w_{+-}(\tau,\vc{u};t,\vc{y})
        \frac{\partial}{\partial u^i}\frac{1}{\nabla^2_u}\partial_\tau\Delta^w_{+-}(\tau,\vc{u};t,\vc{z})
      +\frac{\bar N^2}{6a^2}\frac{\partial}{\partial u^i}\Delta^w_{+-}(\tau,\vc{u};t,\vc{y})
        \frac{\partial}{\partial u^i}\Delta^w_{+-}(\tau,\vc{u};t,\vc{z})
  \bigg]
   + \text{perms}
\nonumber\\
&-&\Delta^w_{-+}(\tau,\vc{u};t,\vc{x})
  \bigg[\frac12\partial_\tau\Delta^w_{-+}(\tau,\vc{u};t,\vc{y})
        \partial_\tau\Delta^w_{-+}(\tau,\vc{u};t,\vc{z})
\label{3point:total}\\
&&\hskip -3.5cm
      +\,\frac13\frac{\partial}{\partial u^i}\partial_\tau\Delta^w_{-+}(\tau,\vc{u};t,\vc{y})
        \frac{\partial}{\partial u^i}\frac{1}{\nabla^2_u}\partial_\tau\Delta^w_{-+}(\tau,\vc{u};t,\vc{z})
      +\frac{\bar N^2}{6a^2}\frac{\partial}{\partial u^i}\Delta^w_{-+}(\tau,\vc{u};t,\vc{y})
        \frac{\partial}{\partial u^i}\Delta^w_{-+}(\tau,\vc{u};t,\vc{z})
  \bigg]
   + \text{perms}
\bigg\}
\,,
\quad
\nonumber
 \eea
where perms denote the six permutations of $\{\vc{x},\vc{y},\vc{z}\}$.
When Eq.~(\ref{3point:total}) is transformed to the spatial Fourier space,
and inflationary mode functions are used for the scalar propagators,
one can calculate the induced bispectrum, which in turn can be related
to the cosmic background temperature fluctuations.
As expected, the expression~(\ref{3point:total})
leads to a manifestly real result  since
$\Delta^w_{+-}=-(\Delta^w_{-+})^*\,$, and coincides with that obtained in the operator
formalism~\cite{Maldacena:2002vr} {\it via}
\beq
\langle\Omega|w(t,\vc{x})w(t,\vc{y})w(t,\vc{z})|\Omega\rangle
 = i\int\limits_{-\infty}^{t}\!\!d^3\vc{u}\,d\tau
  \langle\Omega|\big[w(t,\vc{x})w(t,\vc{y})w(t,\vc{z}),\,
{\cal S}_{\rm I}(\tau,\vc{u})\big]|\Omega\rangle
 \,,
\nonumber
\eeq
where ${\cal S}_{\rm I}$ is the interaction action in the tensor gauge and all fields are taken to
be free Heisenberg fields. Thus, we have reproduced the results
already obtained using the operator formalism and canonical
quantization~\cite{Maldacena:2002vr}.

\subsection{The 4-point Function $\langle w^4 \rangle$}
\label{The 4-point Function}

The quartic interaction terms in (\ref{Swwww}) contribute diagrams of the form
\vskip 0.5cm

\beq\parbox{20mm}{
\begin{fmffile}{quartic1+}
\begin{fmfgraph*}(30,30)
\fmfleftn{i}{2}\fmfrightn{o}{2}
\fmflabel{+}{i1}
\fmflabel{+}{i2}
\fmflabel{+}{o1}
\fmflabel{+}{o2}
\fmf{plain}{i1,v1,i2}
\fmf{plain}{o1,v1,o2}
\fmflabel{+}{v1}
\end{fmfgraph*}
\end{fmffile}}
+\qquad\parbox{20mm}{
\begin{fmffile}{quartic1-}
\begin{fmfgraph*}(30,30)
\fmfleftn{i}{2}\fmfrightn{o}{2}
\fmflabel{+}{i1}
\fmflabel{+}{i2}
\fmflabel{+}{o1}
\fmflabel{+}{o2}
\fmf{plain}{i1,v1,i2}
\fmf{plain}{o1,v1,o2}
\fmflabel{-}{v1}
\end{fmfgraph*}
\end{fmffile}}
\eeq

\vskip 0.5cm

\noindent However, the leading order contribution to the 4-point function comes from (\ref{cubic-aux}), corresponds to diagrams like

\vskip 0.5cm

\beq\label{quartic-aux}
\parbox{20mm}{\begin{fmffile}{aux+}
\begin{fmfgraph*}(30,30)
\fmfleftn{i}{2}\fmfrightn{o}{2}
\fmflabel{+}{i1}
\fmflabel{+}{i2}
\fmflabel{+}{o1}
\fmflabel{+}{o2}
\fmflabel{+}{v1}
\fmflabel{+}{v2}
\fmf{plain}{i1,v1,i2}
\fmf{dashes}{v1,v2}
\fmf{plain}{o1,v2,o2}
\end{fmfgraph*}
\end{fmffile}}
+\qquad
\parbox{20mm}{\begin{fmffile}{aux-}
\begin{fmfgraph*}(30,30)
\fmfleftn{i}{2}\fmfrightn{o}{2}
\fmflabel{+}{i1}
\fmflabel{+}{i2}
\fmflabel{+}{o1}
\fmflabel{+}{o2}
\fmflabel{-}{v1}
\fmflabel{-}{v2}
\fmf{plain}{i1,v1,i2}
\fmf{dashes}{v1,v2}
\fmf{plain}{o1,v2,o2}
\end{fmfgraph*}
\end{fmffile}}
\quad ,\eeq

\vskip 0.5cm

\noindent where the dashed line represents the ``propagation'' of auxiliary fields. As is evident from the form of the auxiliary propagators, such diagrams correspond to effective 4-point interactions.
Since $\Delta_{\pm\mp} = 0$ for all of the auxiliary fields,
there can be no $+-$ internal lines involving auxiliary fields.
The leading order ${\cal O}(\epsilon^1)$ terms from
Eq.~(\ref{cubic-aux}), determining the vertices of the diagrams (\ref{quartic-aux}), read
\bea
 S_{ww(aux)} &=& \int {\rm d}^3\!x{\rm d}t \bar Na^3
\bigg\{
 -\frac{\epsilon}{18}\frac{\tilde n}{\bar N}
    \Big[\dot w^2+\Big(\frac{\partial_iw}{a}\Big)\Big(\frac{\partial_iw}{a}\Big)\Big]
 +\frac{\epsilon}{18}\frac{\nabla^2\tilde S}{(3-\epsilon)\bar Na^2H}
    \Big[\dot w^2+\Big(\frac{\partial_iw}{a}\Big)\Big(\frac{\partial_iw}{a}\Big)\Big]
\nonumber\\
 &&\hskip 2cm
     -\, \frac{\epsilon}{9}\frac{(\tilde N^T_{i}+\partial_i\tilde S)}{\bar Na^2}\dot w(\partial_iw)
\bigg\}
\,.
\label{Swwaux:lo}
\eea

Diagrams like (\ref{quartic-aux}) contribute products of the form,
\bea
 -\frac12\langle\Omega|\hat {\cal O} S_{ww(aux)}S_{ww(aux)}|\Omega\rangle
  &=&-\frac12\int d^3 \vc{u}\int_{-\infty}^{\infty}d\tau \bar N a^3
   \Big\langle\Omega\Big|\hat {\cal O}\bigg\{
    -\frac{\epsilon^2}{81}(\dot w\partial_i w)\frac{1}{\nabla^2}
           (\dot w\partial_iw)
\label{4point:step}
\\
  && \hskip -3cm
    +\, \frac{\epsilon^2(1+\epsilon)}{324}(\dot w\partial_iw)
        \frac{\partial_i\partial_j}{\nabla^4}
           (\dot w\partial_jw)
    + \frac{\epsilon^2}{324H}\Big[(\dot w)^2+\Big(\frac{\partial_jw}{a}\Big)^2\Big]
        \frac{\partial_i}{\nabla^2}(\dot w\partial_iw)
   \bigg\}\Big|\Omega\Big\rangle
\,,
\nonumber
\eea
where $\hat {\cal O}$ is an operator and we used the auxiliary propagators~(\ref{G3}--\ref{G5}). Note
that the contributions from the two auxiliary fields linking the
terms $\tilde n-\nabla^2\tilde S/[(3-\epsilon)a^2H]$ cancel
each other out. We see that the exchange of auxiliary fields induces an effective 4-point interaction given by the expression inside the curly brackets of (\ref{4point:step}). This effective interaction corresponds to the leading order terms of the quartic interaction terms derived in \cite{Seery:2006vu, Jarnhus:2007ia}.

The contribution to the four point function is obtained when
$\hat{\cal O}\rightarrow w_+(t,\vc{x})w_+(t,\vc{y})w_+(t,\vc{z})w_+(t,\vc{v})$. The result is,
\bea
&&\hskip -1cm
\langle\Omega|w(t,\vc{x})w(t,\vc{y})w(t,\vc{z})w(t,\vc{v})|\Omega\rangle
\nonumber\\
  \!&=&\!\int d^3 \vc{u}\int_{-\infty}^{t}d\tau \bar N a^3
   \bigg\{
    \frac{\epsilon^2}{162}\Big(\frac{\partial}{\bar N\partial\tau}\Delta_{+-}^w(\tau,\vc{u};t,\vc{x})
            \frac{\partial}{\partial u^i}\Delta_{+-}^w(\tau,\vc{u};t,\vc{y})\Big)\frac{1}{\nabla_{\vc{u}}^2}
           \Big(\frac{\partial}{\bar N\partial\tau}\Delta_{+-}^w(\tau,\vc{u};t,\vc{z})
            \frac{\partial}{\partial u^i}\Delta_{+-}^w(\tau,\vc{u};t,\vc{v})\Big)
\nonumber
\\
  && \hskip -0cm
    -\, \frac{\epsilon^2(1+\epsilon)}{648}\Big(\frac{\partial}{\bar N\partial\tau}\Delta_{+-}^w(\tau,\vc{u};t,\vc{x})
            \frac{\partial}{\partial u^i}\Delta^w_{+-}(\tau,\vc{u};t,\vc{y})\Big)
        \frac{1}{\nabla_{\vc{u}}^2}\frac{\partial}{\partial u^i}\frac{\partial}{\partial u^j}
           \Big(\frac{\partial}{\bar N\partial\tau}\Delta^w_{+-}(\tau,\vc{u};t,\vc{z})
            \frac{\partial}{\partial u^j}\Delta^w_{+-}(\tau,\vc{u};t,\vc{v})\Big)
\nonumber\\
&&    -\, \frac{\epsilon^2}{648H}\bigg[\Big(\frac{\partial}{\bar N\partial\tau}\Delta^w_{+-}(\tau,\vc{u};t,\vc{x})
            \frac{\partial}{\bar N\partial\tau}\Delta^w_{+-}(\tau,\vc{u};t,\vc{y})\Big)
  +\,\Big(\frac{\partial}{a\partial u^j}\Delta^w_{+-}(\tau,\vc{u};t,\vc{x})\Big)
            \Big(\frac{\partial}{a\partial u^j}\Delta^w_{+-}(\tau,\vc{x};t,\vc{y})\Big)\bigg]
\nonumber\\
  &&\hskip 1.1cm \times\,
\frac{1}{\nabla_{\vc{u}}^2}\frac{\partial}{\partial u^i}
  \Big(\frac{\partial}{\bar N\partial\tau}\Delta^w_{+-}(\tau,\vc{u};t,\vc{z})
            \frac{\partial}{\partial u^i}\Delta^w_{+-}(\tau,\vc{u};t,\vc{v})\Big)-(+-\rightarrow -+)
   + {\rm perms}
   \bigg\}
\,,
\label{4point:final}
\eea
where $perms$ denote the 24 permutations of $\{\vc{x},\vc{y},\vc{z},\vc{v}\}$.
This leading order contribution is of the order ${\cal O}(\epsilon^2)$.
There is also the tree level contribution coming from
the graviton exchange. The leading order vertex in Eq.~(\ref{Swwh})
is of the order ${\cal O}(\epsilon^1)$, and hence formally it also contributes
at the order ${\cal O}(\epsilon^2)$. Here we do not discuss
this contribution any further, and refer the reader
to Ref.~\cite{Seery:2008ax} where the graviton exchange
contribution to the 4-point function has been estimated. Finally,
the terms in Eq.~(\ref{Swwww}) yield contributions that are of
the order ${\cal O}(\epsilon^4)$; since their contribution to the four
point function is suppressed with respect to~(\ref{4point:final}),
we do not calculate them here.

\section{Discussion}
\label{Discussion}

In this paper we developed a path integral formulation for inflationary perturbations in single field inflation. Using a phase space formulation of the system as a starting point, we were able to bring the free action into a particularly simple form which allows for a straightforward calculation of the various propagators, recovering in the process standard results of linear gauge-invariant cosmological perturbation theory. The interaction terms can then be obtained without the necessity of first solving the energy and momentum constraints as has been done so far in the literature. A notable feature is the appearance of two types of auxiliary fields in the path integral beyond the physical propagating degrees of freedom: a set of commuting non-dynamical fields related to the existence of the constraints, as well as a set of anticommuting Faddeev-Popov ghost fields induced by the imposition of gauge conditions. The resulting path integral is independent of the choice of gauge, provided the asymptotic states are defined in terms of the linearized fields and the corresponding linearized gauge transformations.

We then briefly described how to obtain N-point expectation values and commented on the meaning of two different gauges and their relation to cosmological observables in this formalism. They are the uniform field gauge and the tensor gauge which have been widely used in past considerations of non-Gaussianity but in the context of an interaction picture operator approach. Correlators can be expressed in a systematic expansion in diagrams in the in--in formalism. We found
that, when quantum loops are taken into account, anticommuting ghost fields must be included in the computation. Furthermore, internal lines in diagrams should also include the commuting auxiliary fields and we demonstrated their role in the computation of 4-point functions, in which the leading order contributions indeed come from diagrams with internal lines involving auxiliary fields. This way of obtaining the effective 4-point interactions seems simpler than having to go through the solution of the constraints at second order and the substitution of these solutions back in the action~\cite{Seery:2006vu,Jarnhus:2007ia}.

So far, we have only considered standard single field inflation but the generalization to multi-field and more complicated models should be straightforward as long as the canonical formulation is known. We have also explicitly considered only tree diagrams in the tensor gauge, noting that the standard non-linear redefinitions (gauge transformations) can be used to obtain results for the curvature perturbation on comoving slices. Of course we could have obtained this result by working directly in the uniform field gauge, at the price of more complicated interaction terms. Relating results in the two gauges would be particularly interesting if quantum loop corrections are taken into account and questions of renormalization arise. This is a regime where the role of the anticommuting ghosts would become crucial. Currently such questions are only addressed using the (essentially classical) $\Delta N$ formalism on long wavelengths. Other issues involve the appearance of infrared divergences, backreaction and a more rigorous definition of stochastic inflation. A path integral formulation might prove very useful for such considerations which will be the focus of future work.

\section*{Acknowledgements}

\noindent This work was partly initiated during
the ``Non-Linear Cosmological Perturbations" workshop which took place
at the Yukawa Institute for Theoretical Physics (Kyoto) 13-24 April 2009 (YITP-W-09-01) and G.R. wishes to thank the organizers for the hospitality and especially
Misao Sasaki and Takahiro Tanaka for stimulating discussions. G.R. would also like to thank Kari Enqvist for inspiring comments.
During this research, G.R was partly supported by the EU 6th Framework
Marie Curie Research and Training network "UniverseNet" (MRTN-CT-2006-035863).

\section*{Appendices}

In the following appendices we give details on the derivation of
equations (\ref{perturbation hamiltonian})
and~(\ref{interaction hamiltonian}) (Appendices A and B);
the Poisson brackets needed to calculate~(\ref{ghost-operator})
(Appendix C); and the general interaction hamiltonian as well an
intermediate result needed for cubic and quartic vertices
evaluated in section IV (Appendix D).


\subsection*{Appendix A}

Let us consider the spatial Ricci scalar contribution the
action~(\ref{action1}--\ref{Momentum:ADM}).
A spatial metric perturbation $\delta g_{ij} = a^2 h_{ij}$
on a flat FLRW background can be written as
\begin{equation}
  g_{ij} = a^2 (\delta_{ij} + h_{ij})
\,.
\end{equation}
Its matrix inverse $g^{ij} = a^{-2}\tilde g^{ij}$ is then
\begin{equation}
  \tilde g^{ij} = \delta^{ij} - h^{ij} + h^{il}h_l^{\;j}
                - h^{il}h_{l}^{\;k}h_k^{\;j}
                + h^{il}h_{l}^{\;k}h_k^{\;m}h_m^{\;j} + {\cal O}(h^5)
\,,
\label{tilde gij}
\end{equation}
where the indices are raised with the Kronecker delta,
$h^{i}_{\;j} = \delta^{il}h_{lj}$, $h^{ij} = \delta^{il}h_{l}^{\;j}$.
Eq.~(\ref{tilde gij}) can be also written in matrix notation as,
\begin{equation}
  \tilde{\mathbf g} = \mathbf I - \mathbf h + \mathbf h\cdot \mathbf h
                    - \mathbf h\cdot \mathbf h\cdot \mathbf h
                    +- .. +(-1)^n \mathbf h^n
                    + {\cal O}({\mathbf h}^{n+1})
\,.
\label{tilde g:matrix}
\end{equation}
We will find the following notation useful: a numerical subscript $n$ will
indicate that terms of order $n$ and higher in $h_{ij}$ are included.
For example
$\tilde{\mathbf g}_{\geq n}\equiv (-1)^n\mathbf h^n
      + {\cal O}({\mathbf h}^{n+1})$ {\it etc.}
 To calculate the determinant
$\tilde{g}^{\pm 1/2} = [{\rm Det}(\vc{I}+\vc{h})]^{\pm 1/2}$,
we can use
\beq
[{\rm Det}(\vc{I}+\vc{h})]^{\pm 1/2}
= \exp \pm \frac{1}{2} {\rm Tr} \ln (\vc{I}+\vc{h})
 = \sum\limits_{n=0}^\infty\frac{(\pm 1)^n}{2^nn!}
 \left({\rm Tr} \sum\limits_{m=1}^\infty
 \frac{(- 1)^{m-1}}{m}\mathbf h\right)^n
\,.
\label{tilde g:det}
\eeq
For example, in accordance with this notation, we have $(h =  {\rm Tr}[\mathbf h])$
\begin{eqnarray}
(\tilde{g}^{\pm 1/2}\,)_{\geq 2}
&=& \left[\mp\frac14 {\rm Tr} (\mathbf h^2)
    + \frac18 h^2\right]
          + \left[\pm\frac16{\rm Tr}(\mathbf h^3) - \frac18 h {\rm Tr}(\mathbf h^2)
                  \pm \frac{1}{48}h^3\right]
\nonumber\\
 && + \left[\mp\frac18{\rm Tr}(\mathbf h^4) + \frac1{32} {\rm Tr}(\mathbf h\cdot\mathbf h)^2
                  + \frac{1}{12}h {\rm Tr}(\mathbf h^3) \mp \frac{1}{32}h^2{\rm Tr}(\mathbf h^2)
                  + \frac{1}{384}h^4
            \right]+\, {\cal O}(\mathbf h^5)
\,.
\label{root of g}
\end{eqnarray}
We will also be using
\beq
A_{ijkl}\equiv \delta_{ik}\delta_{jl}+\delta_{il}\delta_{jk}
 -\delta_{ij}\delta_{kl}
\,.
\eeq
We can now write the spatial connection as
\begin{equation}
 \Gamma^i_{jl} = \frac12 \tilde g^{ik}
          \left[\partial_jh_{kl}+\partial_lh_{jk}-\partial_kh_{jl}\right]
\,.
\label{connection:spatial}
\end{equation}
 The corresponding Ricci scalar $R = g^{ij}R_{ij}$, with
$R_{ij} = \partial_k \Gamma^k_{ij}-\partial_j \Gamma^k_{ik}
 +\Gamma^l_{kl} \Gamma^k_{ij}-\Gamma^l_{kj} \Gamma^k_{il}$
is then,
\begin{eqnarray}
R &=& \frac{\tilde g^{ij}}{a^2}
    \Bigl\{\frac12(\partial_k \tilde g^{kl})
              \left(\partial_ih_{lj}+\partial_jh_{il}-\partial_lh_{ij}\right)
     + \frac12 \tilde g^{kl}
       \left(\partial_k\partial_ih_{lj}+\partial_k\partial_jh_{il}
             -\partial_k\partial_lh_{ij}\right)
     - \frac12(\partial_j \tilde g^{kl})\left(\partial_ih_{kl}\right)
     - \frac12 \tilde g^{kl}\left(\partial_j\partial_ih_{kl}\right)
\nonumber\\
 &+&\!\!\!\!\frac14\tilde g^{lm} \tilde g^{kn}(\partial_k h_{ml})
              \left(\partial_ih_{nj}+\partial_jh_{in}-\partial_nh_{ij}\right)
     -\frac14\tilde g^{lm} \tilde g^{kn}
      \left(\partial_k h_{mj}+\partial_j h_{km}-\partial_m h_{kj}\right)
      \left(\partial_ih_{nl}+\partial_lh_{in}-\partial_nh_{il}\right)
    \Bigr\}
\,.\;\;\; \label{Ricci scalar:spatial}
\end{eqnarray}
Now making use of
\begin{equation}
 \partial_l\tilde g^{ij} = -\tilde g^{in}(\partial_l h_{nm})\tilde g^{mj}
\,,
\end{equation}
and upon summing various identical terms, Eq.~(\ref{Ricci scalar:spatial})
can be written in the more symmetric form as,
\begin{eqnarray}
\sqrt{g}R &=& a\sqrt{\tilde g}\tilde g^{ij}
    \biggl\{
\tilde g^{kl}[\partial_k\partial_i h_{lj}-\partial_i\partial_j h_{kl}]
 + \tilde g^{km}\tilde g^{ln}
  \Bigl[-(\partial_kh_{mn})(\partial_ih_{lj})
        -\frac14(\partial_lh_{jm})(\partial_ih_{nk})
\nonumber\\
 &&\hskip 1.3cm
        -\frac14(\partial_lh_{jm})(\partial_kh_{in})
        -\frac14(\partial_nh_{ij})(\partial_lh_{km})
        +(\partial_lh_{ij})(\partial_kh_{mn})
        +\frac34(\partial_ih_{kl})(\partial_jh_{mn})
\Bigr]
    \biggr\}
\,,
\label{Ricci scalar:spatial2}
\end{eqnarray}
where (see Eq.~(\ref{root of g}))
\begin{eqnarray}
  \sqrt{g} &=& a^3 \sqrt{\tilde g}
\nonumber\\
   \sqrt{\tilde g} &=& 1 + \frac 12 h
                    +  (\sqrt{\tilde g}\;)_{\geq 2}
\,.
\label{sqrt tilde g}
\end{eqnarray}
Formula~(\ref{Ricci scalar:spatial2})
contains vertices to all orders generated by the $\sqrt{g}R$ term,
and it is written such that it is relatively easy to extract its
contribution to an arbitrary $n$-point vertex. For example, the linear
contribution in $h_{ij}$ is
\begin{eqnarray}
[\sqrt{g}R]_{\rm lin} &=& a
      [\partial_i\partial_j h_{ij}-\partial_i^2 h]
\,,
\label{Ricci scalar:spatial:linear}
\end{eqnarray}
while the quadratic contribution reads,
\begin{eqnarray}
[\sqrt{g}R]_{\rm quad} &=& a \Bigl[
            \frac{h}{2}(\partial_i\partial_jh_{ij}-\nabla^2h)
           - h_{kl}(2\partial_k\partial_i h_{li}-\nabla^2h_{kl})
           + h_{ij}\partial_i\partial_jh
\nonumber\\
      &&   -\,\frac32(\partial_k h_{kl})(\partial_ih_{il})
         -\frac14(\partial_ih)(\partial_i h)
        +(\partial_ih)(\partial_j h_{ji})
        +\frac34 (\partial_ih_{jl})(\partial_i h_{jl})
\Bigr]
\nonumber\\
 &=& a \Bigl[-\frac14h\partial_i^2 h
        +\frac12 h\partial_i\partial_j h_{ij}
        -\frac12 h_{ij}\partial_i\partial_l h_{jl}
        +\frac14 h_{jl}\nabla^2 h_{jl}
\Bigr] + ({\tt tot.\;der.})
\,,
\label{Ricci scalar:spatial:quadratic}
\end{eqnarray}
where in the last line we ignored several total spatial derivative
terms ({\tt tot.\;der.}), which drop out when inserted into the
(free part of the) action~(\ref{action1}).


\subsection*{Appendix B}

We now present steps in the derivation of the free
action~(\ref{2nd-w-h}) and~(\ref{2nd-order}) below.
Let us first analyze the momentum part of the action~(\ref{action1}--\ref{Momentum:ADM})
quadratic in perturbations (see Eq.~(\ref{perturbation hamiltonian})),
which can be written in the form,
\beq
S^{(2)}_{\pi} = \int d^3x\bar{N} dt \bigg\{ -\frac{\mc{P}_\phi^2}{2a^3}
                 \left(\pi_\varphi^2 + I_\varphi\pi_\varphi\right)
              - 4a^3H^2
               \left[\pi^{ij}\frac{A_{ijkl}}{2}\pi^{kl}+ I_{ij}\pi^{ij}\right]
                  \bigg\}
\,,
\label{momentum-action}
\eeq
where
\begin{eqnarray}
I_\varphi &=& -h - \frac{2a^3}{\mc{P}_\phi}\dot\varphi + \frac{2}{\bar{N}}n
\nonumber\\
I_{ij} &=& \frac{1}{2H}\dot h_{ij}
          + 2h_{ij}
          -\frac{\delta_{ij}}{2}h
          -\frac{\delta_{ij}}{\bar{N}}n
           -\frac{1}{\bar N a^2H}\partial_{(i}N_{j)}
\nonumber\\
I &\equiv& \delta_{ij}I_{ij}
     = \frac{1}{2H}\dot h
          + \frac12h
          -\frac{3}{\bar{N}}n-\frac{1}{\bar N a^2 H}\,\nabla^2S
\,.
\label{Iphi-Iij-I}
\end{eqnarray}
 It is now convenient to complete the square in Eq.~(\ref{momentum-action})
 by the appropriate shifts
of the momenta $\pi^{ij}$ and $\pi_\phi$ of the gravitational and scalar
fields, respectively. From Eq.~(\ref{momentum-action})
we easily see that the shifted momenta are,
\bea
\rho^{ij}&=&\pi^{ij}+\frac{1}{2}(I_{ij}- \delta_{ij}I)
\label{mom-shift1}
\\
\rho_\varphi&=&\pi_\varphi+\frac{1}{2}I_\varphi
\,.
\label{mom-shift2}
\eea
With these we get for the momentum terms~(\ref{momentum-action})
\beq
S^{(2)}_\pi =  \int d^3x\bar N dt
      \bigg\{
             -\frac{\mc{P}_\phi^2}{2a^3}\,\,\rho_\varphi^2
     - 4a^3H^2\rho^{ij}\frac{A_{ijkl}}{2}\rho^{kl}
             + \frac{\mc{P}_\phi^2}{8a^3}\,I_\varphi^2
             + a^3H^2(I_{ij}I_{ij}- I^2)
       \bigg\}
\,.
\label{momentum-action2}
\eeq
The linear shifts (\ref{mom-shift1}) and (\ref{mom-shift2}) decouple
the momenta from the inflaton and the metric perturbations.

Next we observe from (\ref{momentum-action2}) that the momenta shifts (\ref{mom-shift1}) and (\ref{mom-shift1}) have introduced terms quadratic in the lapse perturbation $n$. Along with the original linear terms in the
action~~(\ref{S_F}), they read
\begin{equation}
S^{(2)}_{n}  = \int d^3x dt \bigg\{
     \bigg[-\frac{6a^3H^2}{\bar N}+\frac{\mc{P}_\phi^2}{2a^3\bar N}\bigg]n^2
                     +nI_n
                  \bigg\}
\,,
\label{quadratic action:lapse}
\end{equation}
where
\begin{eqnarray}
  I_n = 2a^3H\dot h - \mc{P}_\phi\dot\varphi
       - a^3V_{,\phi}\varphi
       + a (\partial_i\partial_j h_{ij}-\nabla^2 h)
       -\frac{4aH}{\bar N}\nabla^2 S
\,.
\label{In}
\end{eqnarray}
 Just like in the case of the momentum terms above,
we now complete the square for the lapse. The 
contribution is,
\begin{equation}
 S^{(2)}_n = \int d^3x \bar Ndt \bigg\{-\frac{a^3V}{\bar N^2}\tilde n^2
             + \frac{1}{4a^3V}I_n^2\bigg\}
 \,,\qquad \tilde n = n -\frac{\bar N}{2a^3V}I_n
\,.
\label{quadratic action:lapse2}
\end{equation}
%
%
%
%
From the above manipulations we have the shift $N_i$ also contributing to the quadratic action
\begin{equation}
S^{(2)}_{N_i}  = \int d^3x dt \bigg\{
                     \frac{1}{a\bar N}
                 \left([\partial_{(i}N_{j)}]^2  - \alpha (\partial_iN_i)^2
                         + J_{ij}\partial_{(i}N_{j)} \right)
                  \bigg\}
\,,
\label{quadratic action:shift}
\end{equation}
where
\begin{equation}
 \alpha  =
              \frac{1}{3}\Big(1-\frac{\mc{P}_\phi^2}{a^6V}\Big)
\label{alpha}
\end{equation}
and
\begin{eqnarray}
 J_{ij} &=& \bar N a^2
   \bigg[
       -\dot h_{ij}
     +\delta_{ij}
        \bigg(
           \alpha\dot h -\frac{2H}{a^2V}(\partial_k\partial_l h_{kl}-\nabla^2 h)
           +\frac{2H\mc{P}_\phi}{a^3V}\dot\varphi
           +\Big(\frac{\mc{P}_\phi}{a^3}+\frac{2HV_{,\phi}}{V}\Big)\varphi
        \bigg)
   \bigg]
\,,\qquad J=\delta_{ij}J_{ij}
\,,\qquad
\label{Jij}
\end{eqnarray}
where in fact only the vector and scalar parts of $J_{ij}$ in~(\ref{Jij})
contribute.
Analogously to the lapse, the shift contribution to the quadratic action can be written
(up to boundary terms) in the form,
\begin{eqnarray}
   S^{(2)}_{N_i} &=& \int d^3x \bar Ndt
                  \Big\{\frac{1}{a\bar N^2}
                 \Big([\partial_{(i}{\tilde N}^T_{j)}]^2
                  + (1-\alpha)(\nabla^2 {\tilde S})^2 \Big)
                  - \frac{a^3}{4} [\partial_{(i}{\dot h}^T_{j)}]^2
\nonumber\\
           && - \frac{a^3}{4(1-\alpha)V}\Big[
               -\frac13\Big(\dot\phi^2\dot h
               + 2V\nabla^2\tilde h\Big)
               +2\dot \phi H\dot\varphi
               +\frac{4H}{3a^2}\nabla^2(h-\nabla^2\tilde h)
               +\Big(\dot\phi V+2HV_{,\phi}\Big)\varphi\Big]^2
               \Big\}
 \,,
\label{quadratic action:shift2}
\end{eqnarray}
where
\begin{eqnarray}
 \nabla^2 \tilde S &=& \nabla^2 S
          +\frac{1}{6(1-\alpha)}\Big[J-2a^2\bar N\nabla^2 \dot{\tilde h}\Big]
\label{tilde S}
\\
 \partial_{(i}{\tilde N}_{j)}^T
   &=& \partial_{(i}N_{j)}^T -\frac{a^2\bar N}{2} \partial_{(i}{\dot h}^T_{j)}
\,,
\label{tilde NiT}
\end{eqnarray}

When all the terms
from Eqs.~(\ref{momentum-action2}), (\ref{quadratic action:lapse2}),
(\ref{quadratic action:shift2}) and (\ref{perturbation hamiltonian}) are combined,
we get the intermediate, complex expression for the action of
quadratic perturbations in $h_{ij}$ and $\varphi$
(here we drop the contributions from the momenta and
constraints),
\begin{eqnarray}
 S^{(2)}_{h_{ij},\varphi} &=& \int d^3x\bar Ndt \frac{a^3}{4}
  \bigg\{
     - 2H^2\bigg[5{\rm Tr}[\mathbf h\cdot \mathbf h] - \frac32 h^2 \bigg]
     - \frac{1}{a^2}\bigg[h\partial_i^2 h
                        -2 h \partial_i\partial_j h_{ij}
                        +2 h_{ij} \partial_i\partial_l h_{jl}
                        - h_{jl} \partial_i^2 h_{jl}
                    \bigg]
\nonumber\\
    &&\hskip 2cm
      -\,\frac{\dot\phi^2}{2}
           \bigg[{\rm Tr}[\mathbf h\cdot \mathbf h] + \frac12 h^2 \bigg]
       - \frac{1}{2a^2}(\partial_i\varphi)^2
       + V \bigg[{\rm Tr}[\mathbf h\cdot \mathbf h] - \frac12 h^2 \bigg]
       - 2V_{,\phi\phi}\varphi^2
       - 2V_{,\phi}h\varphi
\nonumber\\
    &&\hskip 2cm
       + \, \bigg[{\rm Tr}[\dot{\mathbf h}\cdot \dot{\mathbf h}]
                  - {\dot h}^2 \bigg]
       + 8H \bigg[{\rm Tr}[\mathbf h\cdot  \dot{\mathbf h}]
                - \frac 12 h{\dot h} \bigg]
       + 2H^2 \bigg[8{\rm Tr}[\mathbf h\cdot \mathbf h] - 3 h^2 \bigg]
       + 2 \bigg[{\dot\varphi}^2 + \dot\phi h {\dot\varphi}
                 +\frac14 {\dot\phi}^2 h^2\bigg]
\nonumber\\
    &&\hskip 2cm
       + \frac{1}{V}\bigg[2H{\dot h} - {\dot\phi}{\dot\varphi}
                    + \frac{1}{a^2}(\partial_i\partial_j h_{ij}-\partial_i^2h)
                    -V_{,\phi}\varphi
                    \bigg]^2
\label{action2:intermediate}
\\
    &&\hskip 2cm
       - [\partial_{(i}{\dot h}_{j)}^T\,]^2
       -\frac{1}{4H^2V}
       \bigg[
          2H\dot\phi\dot\varphi
     - \frac23\Big(\frac{\dot\phi^2}{2}\dot h+V\partial_i^2\dot{\tilde h}\Big)
         +\frac43H\frac{\partial_i^2(h-\partial_j^2\tilde h)}{a^2}
         +\Big(\dot\phi V+2HV_{,\phi}\Big)\varphi
       \bigg]^2
 \bigg\}
\nonumber
\,.
\end{eqnarray}
This action can be significantly simplified when expressed in terms of
the transverse traceless tensor $h_{ij}^{TT}$ defined in
Eq.~(\ref{hij:SVT}--\ref{hij:SVT2})
and the gauge invariant combination of scalar matter and scalar gravitational
potentials,
\begin{equation}
  \tilde\varphi = \varphi - z (h-\nabla^2 \tilde h)
\,,\qquad z = \frac{\mc{P}_\phi}{6a^3H}
\,,
\label{SM potential}
\end{equation}
which is also known as the Sasaki-Mukhanov variable.
An important step in the derivation of the quadratic action
is a partial integration of the terms
that contain one time derivative acting on perturbations $\varphi$ and
$h_{ij}$. Furthermore, since vectors are not sourced,
their contribution to~(\ref{action2:intermediate}) cancels out,
and the contributions from scalars and tensors separate.
Following this procedure one eventually arrives at
the quadratic action,
\begin{eqnarray}
 S_{\rm F}  &=& \int d^3x\bar{N}dt a^3
       \bigg\{
             -\frac{\mc{P}_\phi^2}{2a^6}\,\rho_\phi^2
             - 4H^2\rho^{ij}\frac{A_{ijkl}}{2}\rho^{kl}
             -\frac{V}{\bar N^2}\tilde n^2
             + \frac{1}{a^4\bar N^2}
                 \Big(\big[\partial_{(i}{\tilde N}^T_{j)}\,\big]^2
                  + \frac{4H^2}{V}(\nabla^2 {\tilde S})^2 \Big)
\nonumber \\
&&\hskip 2cm +\,\frac{1}{2}\bigg[\dot{\tilde\varphi}^2
                    - \Big(\frac{\nabla\tilde\varphi}{a}\Big)^2
            +\Big(3H\frac{\dot z}{z} + \frac{\ddot z}{z}\Big){\tilde\varphi}^2
               \bigg]
\nonumber\\
&&\hskip 2cm +\,\frac{1}{4}\bigg[(\dot h^{TT}_{ij})^2
                     -  \Big(\frac{\nabla h^{TT}_{ij}}{a}\Big)^2
              \bigg]
      \bigg\}
\,,
\label{2nd-order}
\end{eqnarray}
where
\begin{equation}
3H\frac{\dot z}{z} + \frac{\ddot z}{z}
   = - \frac{\dot{\phi}^2V}{4H^2}
     - \frac{\dot{\phi}}{H}V_{,\phi}
     - V_{,\phi\phi}
\,,
\end{equation}
and the shifts $L_0$, $L_i$ $L_{ij}$ and $L_\varphi$
of the shifted momenta and constraints~(\ref{shift-n}--\ref{shift-rhophi})
in the first line of~(\ref{2nd-order}) can be obtained from
Eqs.~(\ref{Iphi-Iij-I}--\ref{mom-shift1}), (\ref{In})
and~(\ref{tilde S}--\ref{tilde NiT}) and read
\bea\label{L0}
L_0 &=& \frac{\bar{N}}{2a^3V}\left\{2a^3H\dot h - \mc{P}_\phi\dot\varphi
       - a^3V_{,\phi}\varphi
       + a (\partial_i\partial_j h_{ij}-\nabla^2 h)
       +\frac{aV}{6\bar{N}H}\left(J-2\bar{N}a^2\nabla^2\dot{\tilde{h}}\right)\right\} - \frac{2H}{a^2V}\nabla^2 \tilde{S}
\\
       \label{Li}
       L_i&=& \frac{a^2\bar{N}V}{12H^2}\partial_i\dot{\tilde{h}}
             -\frac{V}{24H^2}\frac{\partial_i}{\nabla^2}J
             +\frac{a^2\bar{N}}{2}\dot{h}^T_i
\\
\label{Lij}
L_{ij} &=& -\,\frac{1}{4H}\dot{h}_{ij}- h_{ij}
     +\frac{1}{2\bar{N}a^2H}
              \left(\partial_{(i}L_{j)}+\partial_{(i}\tilde{N}_{j)}\right)
\nonumber\\
      && +\,\frac{\delta_{ij}}{2}\bigg[\frac{1}{2H}\dot{h}+h
            -\frac{2}{\bar{N}}L_0-\frac{1}{\bar{N}a^2H}\partial_kL_k
-\frac{2}{\bar{N}}\tilde{n}
 -\frac{1}{\bar{N}a^2H}\Big(1-4\frac{H^2}{V}\,\Big)\nabla^2\tilde{S}\bigg]
\\
\label{Lvarphi}
L_\varphi &=& \frac{1}{2}h+\frac{a^3}{\mc{P}_\phi}\dot{\varphi}
           -\frac{L_0}{\bar{N}}- \frac{\tilde{n}}{\bar{N}}
           +\frac{2H}{\bar{N}a^2V}\nabla^2\tilde{S}
\,,
\eea
with
%
\beq
J= \bar N a^2
   \bigg[-\frac{{\cal P}_\phi^2}{a^6V}\dot h +3\bigg(
           \frac{2H\mc{P}_\phi}{a^3V}\dot\varphi
           +\Big(\frac{\mc{P}_\phi}{a^3}+\frac{2HV_{,\phi}}{V}\Big)\varphi
        -\frac{2H}{a^2V}(\partial_k\partial_l h_{kl}-\nabla^2 h)\bigg)
   \bigg]
\,.
\label{appendix:J}
\eeq

 We shall now argue that the action~(\ref{2nd-order}) is gauge invariant.
To show this we need to consider how the metric and scalar transform under
infinitesimal coordinate transformations, $x^\mu\rightarrow x^\mu+\xi^\mu(x)$,
under which the metric transforms as, $g_{\mu\nu}\rightarrow g_{\mu\nu}-2\nabla_{(\mu}\xi_{\nu)}$,
$\xi_{\mu}=g_{\mu\nu}\xi^\nu$. To evaluate the covariant derivative in the metric transformation need
the Levi-Civit\`a connection to the zeroth order in perturbations.
The nonvanishing components are,
\beq
 \Gamma^{(0)0}_{00}=\dot{\bar N}
\,;\qquad
 \Gamma^{(0)i}_{0j}=\Gamma^{(0)i}_{j0}=\bar N H\delta_{ij}
\,;\qquad
\Gamma^{(0)0}_{ij}=\frac{a^2H}{\bar N}\delta_{ij}
\,.
\nonumber
\eeq
On the other hand, the metric components to linear
order in perturbations are,
\beq
 g_{00}^{(1)}=-2\bar Nn
\,;\qquad
 g_{0i}^{(1)}=a^2\delta_{ij}N^j=N_i
\,;\qquad
 g_{ij}^{(1)}=a^2h_{ij}
\,,
\eeq
where $h_{ij}$ is decomposed as in Eqs.~(\ref{hij:SVT}--\ref{hij:SVT2}).
These then imply the following linear transformation rules,
\bea
n &\rightarrow& n + \dot \xi_0 - \frac{\dot{\bar N}}{\bar N}\xi_0
\,;\qquad N_i^T\rightarrow N_i^T -\bar N \dot \xi_i^T +2\bar NH\xi_i^T
\,;\qquad S\rightarrow S -\bar N \dot \xi - \xi_0  + 2\bar NH\xi
\nonumber\\
\varphi &\rightarrow& \varphi + \frac{\dot\phi}{\bar N}\xi_0
\,;\qquad\qquad\quad h\rightarrow h -2\frac{\nabla^2}{a^2}\xi +\frac{6H}{\bar N}\xi_0
\,;\qquad\quad h_i^T\rightarrow h_i^T -\frac{2}{a^2}\xi_i^T
\,;\qquad\quad
 h_{ij}^{TT}\rightarrow h_{ij}^{TT}
\,,\quad
\label{gauge transformations}
\eea
where $\xi_i = \xi_i^T + \partial_i \xi$ and $\partial_i\xi_i^T=0$. In particular,
$h-\nabla^2\tilde h \rightarrow h-\nabla^2\tilde h +(6H/\bar N)\xi_0$.
Now, from the transformations~(\ref{gauge transformations}) and
equations~(\ref{SM potential}), (\ref{tilde S}--\ref{tilde NiT}), (\ref{Jij})
and~(\ref{In}--\ref{quadratic action:lapse2})
it follows that $h_{ij}^{TT}$, $\tilde \varphi$, $\tilde n$, $\tilde N_i^T$, $\tilde S$, and also the corresponding terms in the action~(\ref{2nd-order}),
are gauge invariant.

 An important question is whether the whole action~(\ref{2nd-order})
is gauge invariant, also when the momentum terms are included.
That the answer is yes can be argued as follows.
When the Hamilton equations for the momenta, obtained
by varying the action~(\ref{action1}), are solved
in terms of the fields and inserted back into the action~(\ref{action1}),
one obtains the standard Einstein-Hilbert action. This then suggests
that one can define the coordinate transformations for the momenta such
that also the action~(\ref{action1}) becomes covariant. When
this program is carried through at linear order in coordinate
transformations, one gets the transformation rules for the momenta
which render the free action~(\ref{2nd-order}) gauge invariant.
 To see how this works in detail, note first that varying the action
with respect to $\rho_\varphi$ and $\rho^{ij}$ yields
$\rho_\varphi=\pi_\varphi + I_\varphi/2=0$ and
$\rho^{ij}=\pi^{ij}+(I_{ij}-\delta_{ij} I)/2=0$, {\it cf.}
Eqs.~(\ref{mom-shift1}--\ref{mom-shift2}), where
$I_\varphi$, $I_{ij}$ and $I=\delta_{ij}I_{ij}$ are defined in
Eqs.~(\ref{Iphi-Iij-I}). These equations are just the Hamilton equations
of the linearized theory and must be gauge invariant, implying that
$\rho^{ij}$ and $\rho_\varphi$ are gauge invariant.
From the transformation rules
\beq
 I_\varphi\rightarrow I_\varphi +\frac{2}{a^2}\nabla^2\xi
                     + \frac{2V_{,\phi}}{\bar N\dot \phi}\xi_0
\,;\qquad
 I_{ij}\rightarrow I_{ij} -\frac{4}{a^2}\partial_{(i}\xi_{j)}
      +\frac{1}{a^2}\delta_{ij}\nabla^2\xi
      + \frac{1}{\bar NHa^2}\partial_i\partial_j\xi_0
      +\frac{1}{\bar NH}(H^2+\dot H)\delta_{ij}\xi_0
\,
\label{transformation:Ivarphi+Iij}
\eeq
and based on the Hamilton equations for the (unshifted) momenta
$\pi_\varphi=-I_\varphi/2$ and  $\pi^{ij}=-(I_{ij}-\delta_{ij} I)/2$,
we conclude that the momenta transform as,
\bea
 \pi_\varphi &\rightarrow& \pi_\varphi -\frac{\nabla^2}{a^2}\xi
                      - \frac{V_{,\phi}}{\bar N\dot \phi}\xi_0
\,;\qquad
\nonumber\\
 \pi^{ij} &\rightarrow& \pi^{ij}  +\frac{2}{a^2}\partial_{(i}\xi_{j)}
      -\frac{1}{2\bar Na^2H}(\partial_i\partial_j-\delta_{ij}\nabla^2)\xi_0
      +\frac{1}{\bar NH}(H^2+\dot H)\delta_{ij}\xi_0
\,.
\label{transformation:pi_varphi+piij}
\eea
From this one can easily reconstruct how the scalar, vector and tensor
components of
$\pi^{ij}=(\delta_{ij}/3)\pi
          + (\partial_i\partial_j-\delta_{ij}\nabla^2/3)\tilde\pi
          + \partial_{(i}\pi^T_{j)}+\pi^{TT}_{ij}$,
transform:
\bea
 (\pi^{ij})^{TT}&\rightarrow&(\pi^{ij})^{TT}
\,;\qquad \qquad \qquad \,
 (\pi^{i})^{T}\rightarrow(\pi^{i})^{T} + \frac{2}{a^2}\xi_i^T
\nonumber\\
 \tilde\pi&\rightarrow&\tilde\pi + \frac{2}{a^2}\xi -\frac{1}{2\bar Na^2H}\xi_0
\,;\qquad
 \pi\rightarrow \pi + \frac{2}{a^2}\nabla^2\xi
             - \frac{1}{\bar Na^2H}\nabla^2\xi_0
             + \frac{3}{\bar NH}\big[H^2+\dot H\big]\xi_0
\,.
\nonumber
\eea
The implications of these transformation rules for the cosmological
perturbation theory are discussed in section~\ref{The path integral}.

 Next, the action~(\ref{2nd-order}) is time reparametrization
invariant. Thus choosing, for example,
$\bar N=1$ or $\bar N=a$ gives the action in physical
(cosmological) or conformal time, respectively.  One often defines
the Mukhanov variable~\cite{Mukhanov:1990me}
 $v=a\tilde\varphi$. For example, when written in conformal time
($\bar N=a, t\rightarrow \eta$) the lagrangian of the scalar part of the
action~(\ref{2nd-order}) will acquire an additional $[a^{\prime\prime}/a]v^2$
contribution (here a {\it prime} refers to a derivative w.r.t. conformal time $\eta$):
\begin{equation}
 S^{(2)}_{v}  = \int d^3xd\eta
             \frac{1}{2}\bigg[{v^\prime}^2
                    - (\partial_i v)^2
  +\frac{(az)^{\prime\prime}}{az}
 v^2
               \bigg]
\,.
\label{2nd-order:scalar:v}
\end{equation}
Another convenient variable is
\begin{equation}
   w = -\frac{\tilde \varphi}{z}
     = (h-\nabla^2 \tilde h) - \frac{\varphi}{z}
\,,
\label{scalar variable:w}
\end{equation}
in terms of which the scalar part of the quadratic action~(\ref{2nd-order})
 simplifies to,
\begin{equation}
 S^{(2)}_{w}  = \int d^3x\bar Ndt a^3 z^2\frac{1}{2}
                          \Big[{\dot w}^2 - \Big(\frac{\partial_iw}{a}\Big)^2\,
                          \Big]
\,,\qquad z = \frac{\dot\phi}{6H}=\sqrt{\frac{\epsilon}{3}}{\rm sign}[\dot\phi]
\,.
\label{2nd-order:scalar:w}
\end{equation}


 \subsection*{Appendix C}

We now give the expressions for the Poisson brackets of the constraints
$\left\{\mc{Q}_\alpha(\vc{x}), C_\beta(\vc{y})\right\}$, which appear
in the measure of the path integral~(\ref{transition Amplitude3}) and
the constraints $C_\beta=C^{(1)}_\beta+C^{\geq 2}_\beta$
are given in Eqs.~(\ref{C(1)_0}--\ref{C(1)_i})
and~(\ref{C0gt2}--\ref{Cigt2}), with the free and interaction hamiltonian
given in Eqs.~(\ref{perturbation hamiltonian})
and~(\ref{interaction hamiltonian}), respectively.
Before we proceed, we first present
the Poisson brackets between the fields and the constraints,
\bea
 \{h_{ij}(\vc x,t),C_0(\vc y,t),\} &=& -2H(\tilde g^{-1/2})
  \Big\{
     \delta_{ij} +[-h_{ij}+h\delta_{ij}-2\pi^{ij}+\pi\delta_{ij}]
     +[hh_{ij}-h_{il}h_{lj}-4\pi^{il}h_{lj}+\pi h_{ij}
       +\pi^{kl} h_{kl}\delta_{ij}]
\nonumber\\
    &&\hskip 1.95cm +\,[-2h_{il}\pi^{lk}h_{kj}+h_{kl}\pi^{lk}h_{ij}]
  \Big\}\delta(\vc x-\vc y)
\label{PB:hij-C0}
\\
 \{h_{ij}(\vc x,t),C_l(\vc y,t)\} &=& -\frac{2}{a^2}
  \Big\{
     \delta_{l(i}\partial^x_{j)}-\Gamma^l_{ij}
  \Big\}\delta(\vc x-\vc y)
\,;\qquad
 \Gamma^l_{ij} = \tilde g^{lk}
                 \Big(\partial_{(i}h_{j)k}-\frac12\partial_kh_{ij}\Big)
\label{PB:hij-Ci}
\\
 \{\varphi(\vc x,t),C_0(\vc y,t),\}
   &=& -\frac{{\cal P}_\phi}{a^3}
        (\tilde g^{-1/2})(1+\pi_\varphi)\delta(\vc x-\vc y)
\label{PB:varphi-C0}
\\
 \{\varphi(\vc x,t),C_i(\vc y,t),\}
   &=& -\frac{1}{a^2}\tilde g^{ij}\partial_j\varphi\delta(\vc x-\vc y)
\,.
\label{PB:varphi-Ci}
\eea
We are now ready to consider different gauges.

\vspace{0.3cm}

\noindent
a) ``Tensor Gauge'':
$\mc{Q}_0=h\,,\quad \mc{Q}_i=\partial_j\left(h_{ij}-\frac{\delta_{ij}}{3}h\right)$
\bea
\left\{\mc{Q}_0(\vc{x}), C_0(\vc{y})\right\}
 &=& - 2H(\tilde{g}^{-\frac{1}{2}})
    \Big[3 + (2h + \pi) + (-h_{ij}h_{ij}+ h^2 - \pi^{ij}h_{ij} + \pi h)
         + (-2h_{ik}\pi^{kl}h_{li} + \pi^{ij}h_{ij} h)
    \Big]\delta(\vc{x}-\vc{y})
\nonumber
\\
\left\{\mc{Q}_0(\vc{x}), C_i(\vc{y})\right\}
 &=& -\frac{2}{a^2}\Big[\partial_i^x
  - \tilde{g}^{il}\Big(\partial_k h_{lk}
                    -\frac{1}{2}\partial_l h\Big)
            \Big]\delta(\vc{x}-\vc{y})
\label{PB:tensor gauge}
\\
\left\{\mc{Q}_i(\vc{x}), C_0(\vc{y})\right\}
 &=& -2H \tilde{g}^{-\frac{1}{2}}
  \Big\{\!-\!\Big(h_{ij}-\frac{\delta_{ij}}{3}h\Big)
        \!-\!2\Big(\pi^{ij}-\frac{\delta_{ij}}{3}\pi\Big)
       \!+\!(h+\pi)\Big(h_{ij}-\frac{\delta_{ij}}{3}h\Big)
       \!-\!\Big(h_{il}h_{lj}-\frac{\delta_{ij}}{3}h_{kl}h_{kl}\Big)
\nonumber\\
   &&  -\,4\Big(\pi^{il}h_{lj}-\frac{\delta_{ij}}{3}\pi^{kl}h_{kl}\Big)
       \!-\!2\Big(h_{il}\pi^{lr}h_{rj}
                -\frac{\delta_{ij}}{3}h_{kr}\pi^{rs}h_{sk}\Big)
       \!+\!\pi^{kl}h_{kl}\Big(h_{ij}-\frac{\delta_{ij}}{3}h\Big)
  \Big\}(\vc y,t)\,\partial_j^x \delta(\vc{x}-\vc{y})
\nonumber
\\
\left\{\mc{Q}_i(\vc{x}), C_j(\vc{y})\right\}
&=& -\frac{1}{a^2}\Big\{
     \delta_{ij}\nabla_x^2 +\frac{1}{3}\partial_i^x\partial_j^x
  -\tilde{g}^{jk}(2\partial_{(i}h_{l)k}-\partial_kh_{il})(\vc y,t)
         \,\partial_l^x
  +\frac{1}{3}\tilde{g}^{jk}(2\partial_lh_{lk}-\partial_kh)(\vc y,t)\,
          \partial_i^x
      \Big\}\delta(\vc{x}-\vc{y})
\,.
\nonumber
\eea
Since the tensor gauge conditions $Q_\alpha=0$ set
the spatial scalars and vectors to zero, $h=0=\tilde h=h^T_i$,
from the relations~(\ref{PB:tensor gauge}) one easily obtains the ghost operators
$\Omega_{\alpha\beta}= \{\mc{Q}_\alpha,C_\beta\}|_{Q_\alpha=0}$
of Eq.~(\ref{ghost-operator}):
\bea
\Omega_{00}
 &=& - 2H(\tilde{g}^{-\frac{1}{2}})^{TT}
    \Big[3 + \rho+L -h_{ij}^{TT}h_{ij}^{TT}
         - (\rho^{ij}+L_{ij})h_{ij}^{TT}
         -2h_{ik}^{TT}(\rho^{kl}+L_{kl})h_{lj}^{TT}
    \Big]\delta(\vc{x}-\vc{y})
\nonumber
\\
\Omega_{0i}
 &=& -\frac{2}{a^2}\partial_i^x\delta(\vc{x}-\vc{y})
\label{ghosts:tensor gauge}
\\
\Omega_{i0}
 &=& -2H (\tilde{g}^{-\frac{1}{2}})^{TT}
  \Big\{\!-\!h_{ij}^{TT}
        \!-\!2\Big(\rho^{ij}+L_{ij}-\frac{\delta_{ij}}{3}(\rho+L)\Big)
       \!+\!(\rho+L) h_{ij}^{TT}
       \!-\!\Big(h_{il}^{TT}h_{lj}^{TT}
               -\frac{\delta_{ij}}{3}h_{kl}^{TT}h_{kl}^{TT}\Big)
\nonumber\\
   &&  -\,4\,\Big((\rho^{il}+L_{il})h_{lj}^{TT}
           -\frac{\delta_{ij}}{3}(\rho^{kl}+L_{kl})h_{kl}^{TT}\Big)
       \!-\!2\Big(h_{il}^{TT}(\rho^{lr}+L_{lr})h_{rj}^{TT}
       -\frac{\delta_{ij}}{3}h_{kr}^{TT}(\rho^{rs}+L_{rs})h_{sk}^{TT}\Big)
\nonumber\\
   &&
       +\,(\rho^{kl}+L_{kl})h_{kl}^{TT}h_{ij}^{TT}
  \Big\}(\vc y,t)\,\partial_j^x \delta(\vc{x}-\vc{y})
\nonumber
\\
\Omega_{ij}
&=& -\frac{1}{a^2}\Big\{
     \delta_{ij}\nabla_x^2 +\frac{1}{3}\partial_i^x\partial_j^x
  -(\tilde{g}^{jk})^{TT}(\partial_ih_{kl}^{TT}+\partial_lh_{ik}^{TT}
                     -\partial_kh_{il}^{TT})(\vc y,t)
         \,\partial_l^x
      \Big\}\delta(\vc{x}-\vc{y})
\,,
\nonumber
\eea
where $L=L_{ij}\delta_{ij}$ is defined in~(\ref{Lij}), where
$h=0=\tilde h=h_i^T$ and $\varphi\rightarrow -[{\cal
P}_\phi/(6a^3H)]w$ are to be exacted, resulting in the shift
functions given in Eqs.~(\ref{L_0:tensor gauge}--\ref{L:tensor
gauge}) below. Finally, from Eqs.~(\ref{tilde gij}) and~(\ref{root
of g}) we infer
\bea
  (\tilde g^{ij})^{TT} &=& \delta^{ij} - h_{ij}^{TT} + h_{il}^{TT}h_{lj}^{TT}
                - h_{il}^{TT}h_{lk}^{TT}h_{kj}^{TT}
                + h_{il}^{TT}h_{lk}^{TT}h_{km}^{TT}h_{mj}^{TT}
                     + {\cal O}((h_{ij}^{TT})^5)
\nonumber
\\
(\tilde{g}^{-\frac{1}{2}})^{TT} &=& 1 +\frac14h_{ij}^{TT}h_{ij}^{TT}
    -\frac16h_{ij}^{TT}h_{jl}^{TT}h_{li}^{TT}
    + \frac18h_{ij}^{TT}h_{jl}^{TT}h_{lk}^{TT}h_{ki}^{TT}
    + \frac{1}{32}h_{ij}^{TT}h_{ij}^{TT}h_{kl}^{TT}h_{kl}^{TT}
    + {\cal O}\big((h_{ij}^{TT})^5\big)
\,.
\nonumber
\eea

\vspace{0.3cm}

b) ``Uniform Field Gauge'':
$\mc{Q}_0=\varphi\,,\quad
\mc{Q}_i=\partial_j\left(h_{ij}-\frac{\delta_{ij}}{3}h\right)$.

In this gauge the relevant Poisson brackets follow
from Eqs.~(\ref{PB:varphi-C0}--\ref{PB:varphi-Ci}),
\bea
\left\{\mc{Q}_0(\vc{x},t), C_0(\vc{y},t)\right\}
   &=& -\frac{\mc{P}_\phi}{a^3}\tilde{g}^{-\frac{1}{2}}
      (1+\pi_\varphi)\,\delta(\vc{x}-\vc{y})
\nonumber
\\
\left\{\mc{Q}_0(\vc{x},t), C_i(\vc{y},t)\right\}
    &=& -\frac{1}{a^2}\tilde{g}^{ij}(\partial_j\varphi)\,\delta(\vc{x}-\vc{y})
\,,
\label{PB:uniform field gauge}
\eea
and the other Poisson brackets
$\left\{\mc{Q}_i(\vc{x},t), C_\beta(\vc{y},t)\right\} $
are identical to the corresponding tensor gauge
expressions in Eq.~(\ref{PB:tensor gauge}).
In this gauge $\varphi=0=\tilde h=h_i^T$, such that
$h_{ij}=(\delta_{ij}/3)h + h_{ij}^{TT}$, and
the corresponding ghosts operators are,
\bea
\Omega_{00} &=& -\frac{\mc{P}_\phi}{a^3}(\tilde{g}^{-\frac{1}{2}})^{\rm ufg}
      (1+\rho_\varphi+L_\varphi)\,\delta(\vc{x}-\vc{y})
\nonumber
\\
\Omega_{0i} &=& 0
\,,
\label{ghosts:uniform field gauge}
\\
\Omega_{i0}
 &=& -2H (\tilde{g}^{-\frac{1}{2}})^{\rm ufg}
  \Big\{\!-\!h_{ij}^{TT}
        \!-\!2\Big(\rho^{ij}+L_{ij}-\frac{\delta_{ij}}{3}(\rho+L)\Big)
       \!+\!\Big(\frac13h+\rho+L\Big)h_{ij}^{TT}
       \!-\!\Big(h_{il}^{TT}h_{lj}^{TT}
            -\frac{\delta_{ij}}{3}h_{kl}^{TT}h_{kl}^{TT}\Big)
\nonumber\\
   &&\hskip 2cm  -\,\frac{4h}{3}\Big((\rho^{ij}+L_{ij})
               -\frac{\delta_{ij}}{3}(\rho+L)\Big)
 -\,4\Big((\rho^{il}+L_{il})h_{lj}^{TT}
               -\frac{\delta_{ij}}{3}(\rho^{kl}+L_{kl})h_{kl}^{TT}\Big)
\nonumber\\
  && \hskip 2cm
       -2\Big[\Big(\frac{\delta_{il}}{3}h+h_{il}^{TT}\Big)
                      (\rho^{lr}+L_{lr})
              \Big(\frac{\delta_{rj}}{3}h+h_{rj}^{TT}\Big)
    -\frac{\delta_{ij}}{3}\Big(\frac{\delta_{kr}}{3}h+h_{kr}^{TT}\Big)
              (\rho^{rs}+L_{rs})
            \Big(\frac{\delta_{sk}}{3}h+h_{sk}^{TT}\Big)\Big]
\nonumber\\
  && \hskip 2cm
        +\,\Big[\frac13(\rho+L)h
          +(\rho^{kl}+L_{kl})h_{kl}^{TT}\Big]h_{ij}^{TT}
  \Big\}(\vc y,t)\,\partial_j^x \delta(\vc{x}-\vc{y})
\nonumber
\\
\Omega_{ij}
&=& -\frac{1}{a^2}\Big\{
     \delta_{ij}\nabla_x^2 +\frac{1}{3}\partial_i^x\partial_j^x
  -\frac13(\tilde{g}^{jk})^{\rm ufg}(\partial_{i}h\delta_{kl}
                     +\delta_{ki}\partial_{l}h
                   -\partial_kh\delta_{il})(\vc y,t)\partial_l^x
\nonumber\\
&&\hskip 0.5cm
  -(\tilde{g}^{jk})^{\rm ufg}(\partial_{i}h_{lk}^{TT}+\partial_{l}h_{ik}^{TT}
                   -\partial_kh_{il}^{TT})(\vc y,t)
         \,\partial_l^x
 -\,\frac{1}{9}[(\tilde{g}^{jk})^{\rm ufg}\partial_kh](\vc y,t)\,
          \partial_i^x
      \Big\}\delta(\vc{x}-\vc{y})
\,,
\nonumber
\eea
where the superscript in $(\tilde g^{-1/2})^{\rm ufg}$ and
 $(\tilde g^{ij})^{\rm ufg}$ signify that only
$h$ and $h_{ij}^{TT}$ contribute to $h_{ij}$, {\it i.e.}
$h_{ij}=(\delta_{ij}/3)h+h_{ij}^{TT}$.

\subsection*{Appendix D: Interaction Hamiltonian, cubic and quartic action}

 The interaction hamiltonian containing cubic and higher order interactions
as inferred from Eqs.~(\ref{ADM:hamiltonian}) and~(\ref{Ricci
scalar:spatial2}) is of the general form:
\bea
\mc{H}_{\rm I}&=& 4\bar{N}a^3H^2
   \Bigg[-\frac32 (\tilde g^{-\frac12})_{\geq 3}
         - (\tilde{g}^{-\frac12})_{\geq 2} h
         + h_{ij} (\tilde{g}^{-\frac12})_{\geq 1}\frac{1}{2}A_{ijkl}h_{kl}
\nonumber\\
&& \hskip 0.7cm
         +\, \pi^{ij}\left(-(\tilde{g}^{-\frac12})_{\geq 2} \delta_{ij}
         + (\tilde{g}^{-\frac12})_{\geq 1}\left(h_{ij}
         - \delta_{ij}h\right)
    + \tilde{g}^{-\frac12}
         \left(h_{ik}h_{kj}-hh_{ij}\right)\right)  \nonumber \\
&& \hskip .7cm
     + \,\, \pi^{ij}\left(
                  (\tilde{g}^{-\frac12})_{\geq 1}\frac{1}{2}A_{ijkl}
          + \tilde{g}^{-\frac12}\left(2h_{jl}\delta_{ik} - \delta_{ij}h_{kl}
                                 + h_{ik}h_{jl}-\frac{1}{2}h_{ij}h_{kl}
                             \right) \right)\pi^{kl}\Bigg]
\nonumber\\
 &&+\,\bar{N}a^3\left( (\tilde{g}^{\frac12})_{\geq 2}
             \sum\limits_{n=1}^\infty \frac{V^{(n)}}{n!}\varphi^n
            + \frac{1}{2}h \sum\limits_{n=2}^\infty \frac{V^{(n)}}{n!}\varphi^n
            + \sum\limits_{n=3}^\infty \frac{V^{(n)}}{n!}\varphi^n
            +  (\tilde{g}^{\frac12})_{\geq 3}V
                \right)
\nonumber\\
 &&+\, \bar{N}\frac{\mc{P}_\phi^2}{2a^3}
           \left((\tilde{g}^{-\frac12})_{\geq 1}\pi_\varphi^2
                + 2(\tilde{g}^{-\frac12})_{\geq 2}\pi_\varphi
               + (\tilde{g}^{-\frac12})_{\geq 3}
           \right)
\nonumber\\
&&+\,\, \frac{a\bar{N}}{2}
           \left((\tilde{g}^{\frac12})_{\geq 1}\delta^{ij}
                 + \tilde{g}^{ij}_{\geq 1}
                 + (\tilde{g}^{\frac12})_{\geq 1}(\tilde{g})_{\geq 1}^{ij}
           \right)\partial_i\varphi\partial_j\varphi
\nonumber\\
&&-\,a\bar{N}\left((\tilde{g}^{\frac12})_{\geq 1}
                 \left((\tilde{g}^{ij})_{\geq 1}\delta^{kl}
                       +\delta^{ij}(\tilde{g}^{kl})_{\geq 1}
                 \right)
 + \tilde{g}^{\frac12}(\tilde{g}^{ij})_{\geq 1}(\tilde{g}^{kl})_{\geq 1}\right)
           \left[\partial_k\partial_i h_{lj}-\partial_i\partial_jh_{kl}\right]
 \nonumber\\
&&-\, a\bar{N} \left(\tilde{g}^{\frac12}\tilde{g}^{ij}\tilde g^{km}\tilde g^{ln}
       -\delta^{ij}\delta^{km}\delta^{ln}\right)
  \Bigl[-(\partial_kh_{mn})(\partial_ih_{lj})
        -\frac14(\partial_lh_{jm})(\partial_ih_{nk})
\nonumber\\
 &&\hskip 2.cm
        -\frac14(\partial_lh_{jm})(\partial_kh_{in})
        -\frac14(\partial_nh_{ij})(\partial_lh_{km})
        +(\partial_lh_{ij})(\partial_kh_{mn})
        +\frac34(\partial_ih_{kl})(\partial_jh_{mn})
\Bigr]
\,,
\label{interaction hamiltonian}
\eea
where $\tilde g_{ij}^{(\geq n)}$ and $(\tilde g^{\pm1/2})^{(\geq
n)}$ are given in Eqs.~(\ref{tilde gij}) and~(\ref{root of g}),
respectively. The cubic part of the
interaction action~(\ref{S_I}) is of the form,
\beq {\cal S}_{\rm cubic} =\int d^3xdt \Big\{-{\cal H}_{\rm
cubic}+nC_0^{(2)}+N_iC_i^{(2)}\Big\} \,, \label{action:cubic}
\eeq
where $C_0^{(2)}$ and $C_i^{(2)}$ are the quadratic part of Eqs.~(\ref{C0gt2}-\ref{Cigt2})
and~(\ref{perturbation hamiltonian}).
When written in tensor gauge~(\ref{gauge-cond1}), the cubic part
of the interaction hamiltonian density~(\ref{interaction
hamiltonian}) reads
\bea - {\cal H}_{\rm cubic} &=& \bar N a^3 \bigg\{2H^2\Big[
     -\frac{3-\epsilon}{3}h_{ij}^{TT}h_{jl}^{TT}h_{li}^{TT}
     +\frac12(\rho+L)h_{ij}^{TT}h_{ij}^{TT}
     -2h_{ij}^{TT}(\rho^{jl}+L_{jl})h_{li}^{TT}
\nonumber\\
 &&  \hskip 1.6cm
     -\,4(\rho^{ij}+L_{ij})h_{jl}^{TT}(\rho^{li}+L_{li})
     + 2(\rho+L)(\rho^{ij}+L_{ij})h_{ij}^{TT}
     -\frac{\epsilon}{2}(\rho_\varphi+L_\varphi)h_{ij}^{TT}h_{ij}^{TT}
\Big]
\nonumber\\
  && \hskip .6cm +\,\frac{h_{ij}^{TT}}{a^2}\Big[
     \frac{1}{4}(\partial_ih_{kl}^{TT})(\partial_jh_{kl}^{TT})
   + \frac{1}{2}(\partial_lh_{jk}^{TT})(\partial_kh_{il}^{TT})
   - \frac{3}{2}(\partial_lh_{ik}^{TT})(\partial_lh_{jk}^{TT})
   \Big]
\nonumber\\
  && \hskip .6cm +\,\frac{\epsilon}{18a^2}h_{ij}^{TT}(\partial_iw)(\partial_jw)
    +\frac{(3-\eta)\epsilon H^2}{6}h_{ij}^{TT}h_{ij}^{TT}w
    +\frac{\epsilon^{3/2}V_{,\phi\phi\phi}}{162}w^3
 \bigg\} \,, \label{H:cubic} \eea
while the cubic constraint contributions read,
\bea
 n C_0^{(2)} &=& \Big(\tilde n + L_0\Big)a^3\Big\{
    -4H^2(\rho^{ij}+L_{ij})(\rho^{ij}+L_{ij})
    +2H^2(\rho+L)^2
    -4H^2(\rho^{ij}+L_{ij})h_{ij}^{TT}
\label{N0C0:cubic}\\
&& \hskip -0.3cm -\,2H^2\epsilon(\rho_\varphi+L_\varphi)^2
    - \frac{\epsilon}{18a^2}(\partial_iw)^2
    -\frac{\epsilon V_{,\phi\phi}}{18}w^2
    +\frac{1}{a^2}h_{ij}^{TT}\nabla^2h_{ij}^{TT}
    +\frac{3}{4a^2}(\partial_lh_{ij}^{TT})(\partial_lh_{ij}^{TT})
    -H^2(1+\epsilon)h_{ij}^{TT}h_{ij}^{TT}
 \Big\}
\nonumber
\\
 N_i C_i^{(2)} &=& (\tilde N_i + L_i)a^3\Big\{
    \frac{2\epsilon H}{3a^2}(\rho_\varphi+L_\varphi)\partial_iw
    -\frac{2\epsilon H}{3a^2}h_{ij}^{TT}\partial_jw
    -\frac{2H}{a^2}(\rho^{lj}+L_{lj})(2\partial_jh_{il}^{TT}-\partial_ih_{jl}^{TT})
 \Big\}
\label{NiCi:cubic}
\,.
 \eea
The shift functions~(\ref{L0}--\ref{appendix:J}) are,
\bea
 L_0 &=& -\frac{\bar N\epsilon}{6}w -\frac{\nabla^2\tilde S}{(3-\epsilon)a^2H}
\label{L_0:tensor gauge}
\\
 L_i &=& \frac{a^2\bar N\epsilon}{6}\frac{\partial_i}{\nabla^2}\dot w
\label{L_i:tensor gauge}\\
 L_\varphi &=& -\frac{\dot w}{6H}+\frac{\eta w}{6}-\frac{\tilde n}{\bar N}
            + \frac{\nabla^2\tilde S}{(3-\epsilon)\bar N a^2H}
 \label{L_varphi:tensor gauge}\\
L_{ij} &=& -\frac{\dot h_{ij}^{TT}}{4H}-h_{ij}^{TT}
-\frac{\epsilon}{12H}\Big(\delta_{ij}-\frac{\partial_i\partial_j}{\nabla^2}\Big)\dot
w  + \frac{\epsilon\delta_{ij}}{6}w +\frac{\partial_{(i}\tilde
N_{j)}^T+\partial_i\partial_j\tilde S}{2\bar
Na^2H}-\frac{\delta_{ij} \tilde n}{\bar
N}-\frac{(1-\epsilon)\delta_{ij}\nabla^2\tilde
S}{2(3-\epsilon)\bar Na^2H}
 \label{L_ij:tensor gauge}\\
L &=& \delta_{ij} L_{ij} = - \frac{\epsilon}{6H}\dot w  +
\frac{\epsilon}{2}w -\frac{3\tilde n}{\bar N}+\frac{\nabla^2\tilde
S}{(3-\epsilon)\bar Na^2H} \,.
 \label{L:tensor gauge}
 \eea
Finally, for the calculation of four point functions, the quartic
contribution from the interaction action~(\ref{S_I}) is needed,
\beq {\cal S}_{\rm quartic} =\int d^3xdt \Big\{-{\cal H}_{\rm
quartic}+nC_0^{(3)}+N_iC_i^{(3)}\Big\} \,, \label{action:quartic}
\eeq
where $C_0^{(3)}$ and $C_i^{(3)}$ are the cubic parts of
Eqs.~(\ref{C0gt2}-\ref{Cigt2}). In tensor gauge they are,
\bea
 nC_0^{(3)} &=& -\Big(\frac{\tilde n}{\bar N}+\frac{L_0}{\bar N}\Big)
                  {\cal H}_{\rm cubic}
\label{nC0:quartic}
\\
 N_iC_i^{(3)}\! &=&\! \big(\tilde N_i\!+\!L_i)a^3
             \bigg\{
               \!-\! \frac{2\epsilon H}{3a^2}(\rho_\varphi+L_\varphi)h_{ij}^{TT}\partial_jw
               + \frac{2\epsilon H}{3a^2}h_{il}^{TT}h_{lj}^{TT}\partial_jw
               +
               \frac{2H}{a^2}h_{im}^{TT}(\rho^{lj}\!+\!L_{lj})(2\partial_jh_{ml}^{TT}
                                  - \partial_mh_{lj}^{TT})
             \bigg\}
\,,\qquad\;
\label{NiCi:quartic}
\eea
while the quartic part of the hamiltonian~(\ref{interaction
hamiltonian}) becomes,
\bea
 -{\cal H}_{\rm quartic} &=& \bar Na^3
 \Bigg\{
    \frac{3}{4}H^2\Big[h_{ij}^{TT}h_{jl}^{TT}h_{lk}^{TT}h_{ki}^{TT}
                  +\frac14h_{ij}^{TT}h_{ij}^{TT}h_{kl}^{TT}h_{kl}^{TT}
               \Big]
    -\frac23H^2(\rho+L)h_{ij}^{TT}h_{jl}^{TT}h_{li}^{TT}
\nonumber\\
  &&\hskip 1cm
    +\,H^2[-h_{ij}^{TT}h_{ij}^{TT}
    -(\rho^{ij}+L_{ij})h_{ij}^{TT}
    -(\rho^{ij}+L_{ij})(\rho^{ij}+L_{ij})
    +\frac12(\rho+L)^2]h_{kl}^{TT}h_{kl}^{TT}
\nonumber\\
  &&\hskip 1cm
    +\,2H^2[-2(\rho^{ij}+L_{ij})h_{jk}^{TT}(\rho^{kl}+L_{kl})h_{li}^{TT}
    +(\rho^{ij}+L_{ij})h_{ij}^{TT}(\rho^{kl}+L_{kl})h_{kl}^{TT}]
\nonumber\\
  &&\hskip 1cm
    +\,\frac{1}{a^2}[-h_{mn}^{TT}h_{mn}^{TT}(\delta_{ij}h_{kl}^{TT}+h_{ij}^{TT}\delta_{kl})
                   -h_{in}^{TT}h_{nj}^{TT}h_{kl}^{TT}-h_{ij}^{TT}h_{kn}^{TT}h_{nl}^{TT}]
       [\partial_k\partial_ih_{lj}^{TT}-\partial_i\partial_jh_{kl}^{TT}]
\nonumber\\
  &&\hskip 0cm
    +\,\frac{1}{a^2}\Big[\delta_{ij}\delta_{km}h_{lp}^{TT}h_{pn}^{TT}
                  +\delta_{ij}h_{kp}^{TT}h_{pm}^{TT}\delta_{ln}
                  +h_{ip}^{TT}h_{pj}^{TT}\delta_{km}\delta_{ln}
                  +\delta_{ij}h_{km}^{TT}h_{nl}^{TT}
                  +h_{ij}^{TT}\delta_{km}h_{nl}^{TT}
\nonumber\\
  &&\hskip 1cm
                  +\,h_{ij}^{TT}h_{km}^{TT}\delta_{nl}
                  -\frac14h_{pr}^{TT}h_{pr}^{TT}\delta_{ij}\delta_{km}\delta_{ln}
                   \Big]
\nonumber\\
  &&\hskip 0cm
   \times\,\Big[\partial_kh_{mn}^{TT}\partial_ih_{lj}^{TT}
          -\frac14 \partial_lh_{jm}^{TT}\partial_ih_{nk}^{TT}
          -\frac14 \partial_lh_{jm}^{TT}\partial_kh_{in}^{TT}
          -\frac14 \partial_nh_{ij}^{TT}\partial_lh_{km}^{TT}
          + \partial_lh_{ij}^{TT}\partial_kh_{mn}^{TT}
          +\frac34 \partial_ih_{kl}^{TT}\partial_jh_{mn}^{TT}
       \Big]
\nonumber\\
  &&\hskip 0cm
  -\,2\epsilon H^2\Big[\frac14(\rho_\varphi+L_\varphi)^2 h_{ij}^{TT}h_{ij}^{TT}
                  -\frac13(\rho_\varphi+L_\varphi)h_{ij}^{TT}h_{jl}^{TT}h_{li}^{TT}
                  + \frac18h_{ij}^{TT}h_{jl}^{TT}h_{lk}^{TT}h_{ki}^{TT}
                  + \frac{1}{32}h_{ij}^{TT}h_{ij}^{TT}h_{kl}^{TT}h_{kl}^{TT}
                \Big]
\nonumber\\
  &&\hskip 0cm
  +\,\frac{\epsilon}{18}\Big[\frac14h_{kl}^{TT}h_{kl}^{TT}\delta_{ij}-h_{il}^{TT}h_{lj}^{TT}
                \Big]\Big(\frac{\partial_iw}{a}\Big)\Big(\frac{\partial_jw}{a}\Big)
\nonumber\\
  &&\hskip 0cm
  +\,\bigg[
      - \frac{\epsilon^2V_{,\phi\phi\phi\phi}}{1944}w^4
      + \frac{\epsilon V_{,\phi\phi}}{72}w^2h_{ij}^{TT}h_{ij}^{TT}
      - \frac{(3-\eta)\epsilon H^2}{9}wh_{ij}^{TT}h_{jl}^{TT}h_{li}^{TT}
 \nonumber\\
  &&\hskip 1cm     +\, (3\!-\!\epsilon)H^2\Big(\frac14h_{ij}^{TT}h_{jk}^{TT}h_{kl}^{TT}h_{li}^{TT}
      -\frac{1}{16}h_{ij}^{TT}h_{ij}^{TT}h_{kl}^{TT}h_{kl}^{TT}\Big)
  \bigg]
 \Bigg\}
\,.
\label{quartic hamiltonian}
\eea
These formulae are used in section IV to construct some of the
cubic and quartic vertices of the theory.

\end{document}